# Empowering Patients Using Smart Mobile Health Platforms: Evidence From A Randomized Field Experiment[1]

Anindya Ghose
Stern School of Business
New York University

Xitong Guo
School of Management
Harbin Institute of Technology

Beibei Li
Heinz College
Carnegie Mellon University

Yuanyuan Dang[2]
School of Management
Harbin Institute of Technology

**Abstract**

With today's technological advancements, mobile phones and wearable devices have become extensions of an increasingly diffused and smart digital infrastructure. In this paper, we examine mobile health (mHealth) platforms and their health and economic impacts on the outcomes of chronic disease patients. To do so, we partnered with a major mHealth firm that provides one of the largest mobile health app platforms in Asia specializing in diabetes care. We designed and implemented a randomized field experiment based on detailed patient health activities (e.g., steps, exercises, sleep, food intake) and blood glucose values from 1,070 diabetes patients over several months. Our main findings show that the adoption of the mHealth app leads to an improvement in health behavior, which in turn leads to both short term metrics (such as reduction in patients' blood glucose and glycated hemoglobin levels) and longer-term metrics (such as hospital visits and medical expenses). Patients who adopted the mHealth app undertook higher levels of exercise, consumed healthier food with lower calories, walked more steps and slept for longer times on a daily basis. They also were more likely to substitute offline visits with telehealth. A comparison of mobile vs. PC enabled version of the same app demonstrates that the mobile has a stronger effect than PCs in helping patients make these behavioral modifications with respect to diet, exercise and lifestyle, which ultimately leads to an improvement in their healthcare outcomes. We also compared outcomes when the platform facilitates personalized health reminders to patients vis-à-vis generic (non-personalized) reminders. Surprisingly, we found that personalized mobile messages with patient-specific guidance can have an inadvertent (smaller) effect on patient app engagement and lifestyle changes, leading to a lower health improvement. However, they are more like to encourage a substitution of offline visits by telehealth. Overall, our findings indicate the massive potential of mHealth technologies and platform design in achieving better healthcare outcomes.

[1] We thank seminar participants in Harvard University, University of Washington, University of British Columbia, University of Georgia, Dartmouth College, CMU-Pitt Seminar series, Boston University, and conference participants at MIT CODE for their helpful comments that have improved the paper. Anindya Ghose and Beibei Li thank the National Science Foundation for supporting this research through an NSF-EAGER grant.
[2] To whom correspondence may be addressed: Xitong Guo (xitongguo@hit.edu.cn), Beibei Li (beibeili@andrew.cmu).

# 1. Introduction

Facilitated by emerging smart mobile health (mHealth) technologies, the healthcare ecosystem has been undergoing a disruptive, digital transformation in transitioning from reactive care to proactive and preventive care that can potentially be administered more cost-effectively. As defined by Estrin and Sim (2010), mHealth is the combination of mobile computing, medical sensor, and communications technologies used for healthcare services, including chronic-disease management and wellness. mHealth includes medical applications that may run on smartphones, tablets, sensors that track vital signs and health activities, and cloud-based computing systems for collecting health data. Indeed, mHealth technologies have demonstrated tremendous potential in shaping the healthcare industry toward a new era of evidence-based medicine and "Quantified Self" (QS)—individuals engaged in the self-tracking of biological, physical, behavioral, and environmental information (e.g., McKinsey 2013). The global mHealth market will reach $49 billion by 2020, growing at a rate of more than 47% between 2013 and 2020.[3]

Given the importance of health behaviors to well-being, health outcomes, and disease processes, mHealth technologies have great potential in facilitating patient life style and behavior modification through patient education, improved autonomous self-regulation and perceived competence. The accessibility, convenience, and ubiquity inherent to mobile devices can help patients easily upload information on a regular basis and follow the guidance that would eventually lead to improved health conditions. However, although there is tremendous promise, uncertainty exists regarding whether mHealth can indeed improve patient health and behavior outcome, for a number of reasons.

First, although mHealth technologies can facilitate easy medical communication and interventions for patients, too frequent interventions might lead to annoyingness or habituation (Pop-

---

[3] http://www.grandviewresearch.com/industry-analysis/mhealth-market

Eleches et al. 2011). Second, health information that is inconsistent with patients' prior belief or perceived as non-credible may be less persuasive and lead to potential information avoidance (Klein and Stefanek 2007). Third, the increased pervasiveness of personal behavioral tracking may bring potential privacy concerns to the users. Previous studies show patients might perceive highly personalized mobile SMS messages as intrusive (e.g., Pop-Eleches et al. 2011).

Furthermore, frequent personalized messages can also cause patients to feel pressured or coerced. Such perceived control and judgment can significantly demotivate patient behavior and lead to lower level of engagement and healthy activities. In particular, prior theories on self-determination (Self-Determination Theory (SDT)) and cognitive evaluation (Cognitive Evaluation Theory (CET)) have demonstrated that lack of choiceful and volitional feeling can lead to loss of autonomy and self-motivation (Deci and Ryan 1985). Such loss of autonomy can lead to lower patient engagement in the health self-management process (e.g., lower mHealth app usage, lower patient-physician engagement, lower compliance to medication and treatment). Also, prior research showed that pressured evaluations and imposed goals diminish intrinsic motivation because they conduce toward an external perceived locus of causality (Deci & Ryan, 1985). Therefore, frequent and advanced personalization enabled by mHealth technologies might backfire patients' engagement in health self-management.

Finally, from a methodological perspective, measuring the effectiveness of mHealth technology on patient health and behavior outcomes can be rather challenging. To date, very little knowledge has been developed toward evaluating the effectiveness of the mHealth applications. Archival analyses using secondary data may not work due to the potential patient self-selection bias in mHealth technology adoption, as well as patient heterogeneity and high dropout rate in mHealth technology usage. Hence, these issues call for a scientific, rigorous approach to evaluate and quantify the effectiveness of the mHealth platforms.

The above challenges motivate us to ask the following research questions in this paper: In the context of healthcare, does the adoption of mHealth technologies (such as mobile apps) persuade patients with chronic diseases to make behavioral modification in their wellness and lifestyle? If so, what is the corresponding impact on patients' healthcare outcomes, in both short-term metrics (such as reduction in patients' blood glucose and glycated hemoglobin levels) and longer-term metrics (such as hospital visits and medical expenses).*?*

Furthermore, to disentangle the underlying mechanism that drives the observed health outcome, we also look into the detailed patient activities, such as daily walking steps, exercise time, sleeping pattern and food intake, documented through the app, together with the detailed app usage data. This enables us to understand how patients actually use the mHealth app, what kinds of behavioral modifications occur in response to the app usage, and the underlying mechanism. Note that looking into the detailed patient activities and app usage logs can help us understand not only *whether or not*, but also *why* mHealth technologies can help improve the healthcare outcomes over time. This is one unique aspect of our work, which distinguishes it from the existing work in this area.

To achieve our goal, in this paper, we instantiate our study within the context of mHealth application for diabetes care. Diabetes is a chronic illness with significant health consequences that lead to macro- and microvascular complications, including heart disease, stroke, hypertension, nephropathy, and neuropathy. The American Diabetes Association (ADA) estimates that 25.8 million children and adults in the United States in 2011 had type 1 or type 2 diabetes. Diabetes poses a heavy economic burden on the US health care system, with estimated associated costs in 2017 of $327 billion (American Diabetes Association 2018). Worldwide, high blood glucose kills about 3.4 million people annually. WHO projects diabetes deaths will double between 2005 and 2030.[4] Therefore, proper patient education and self-management are pivotal, especially for those who are unable to adhere to the

---
[4] http://www.euro.who.int/en/health-topics/noncommunicable-diseases/diabetes/data-and-statistics

complex treatment regimen. However, self-management tasks such as regular medication, frequent blood sugar checks, strict diet management, and consistent exercise can be quite challenging. Hence, there is potential for mHealth applications, to help improve patients' adherence to these behaviors through long-term engagement. Beyond diabetes care, our methodologies and insights have the potential to be generalized to other chronic disease or wellness contexts.

In particular, to evaluate the effectiveness of mHealth applications on diabetes patients' behavior and health outcomes, we partnered with a major mHealth company in Asia that provides the nation's largest mHealth app platform that specializes in diabetes care. We designed and implemented a randomized field experiment based on 9,251 unique observations on blood glucose values and 55,359 unique observations on detailed patient health activities (e.g., steps, exercises, sleep, food intake) and app usage logs from 1,070 diabetes patients over three months together with a follow-up survey after five months. We recruited our participants on a rolling basis. The entire study spanned from May 1, 2015, to July 31, 2016. By randomly assigning patients to different groups (e.g., adoption vs. no adoption of mHealth application), we are able to measure the treatment effect from a causal perspective. Moreover, to evaluate the potential economic impact of the mHealth platform on patients' medical costs and hospital visits, we conducted additional surveys and telephone interviews before and after the experimental period.

Our main findings are as follows. *First*, the adoption of the mHealth platform demonstrates a statistically significant impact on reducing the blood glucose and glycated hemoglobin levels[5] of diabetes patients over time. The mHealth platform also has a statistically significant impact on reducing hospital visits and medical expenses for diabetes patients through increased usage of telehealth.

---

[5] Glycated hemoglobin is a form of hemoglobin that is measured primarily to identify the three-month average blood glucose concentration. Whereas blood glucose is a real-time measurement of blood sugar level. In diabetes research, both measurements are commonly used in evaluating both long-term and real-time blood sugar levels. The reference healthy range for Glucose is recommended to be between 3.9 and 7.1 mmol/L (Millimoles Per Litre). In 2010, the American Diabetes Association Standards of Medical Care in Diabetes also recommended Glycated hemoglobin to be lower than 6.5 mmol/L.

*Second*, the mHealth platform shows a 21.6% stronger impact on patients' health outcome than does the web-based platform (i.e., PC version of the application) that provides the same functions for diabetes management. This finding builds on the prior literature on the differences between PC and mobile devices (e.g., Xu et al. 2016, Ghose 2017), indicating an edge that mobile devices have over PC in affecting patients' health behavior because mobility allows a user to respond more flexibly to real-time information (Ghose et al. 2013).

*Third*, the mHealth platform also demonstrates a significantly stronger impact on patients' dietary and life style improvement as well as engagement with app usage than does the web-based platform. Interestingly, we found that patients who adopted the mHealth application did significantly higher level of daily exercise, consumed healthier food with lower daily calories intake, walked more steps and slept for longer time a day. In particular, when patients double their time of mHealth app usage, we observe an average of 17.1% decrease in food calorie consumption, 5.4% increase in daily exercise time, and 14.5% increase in daily sleeping time, leading to an average of 0.29 mmol/L decrease in blood glucose. Our results suggest that mHealth technology can help patients become more autonomously self-regulated with their health behavior. Such increasing intrinsic motivation can help patients become more engaged, persistent and stable in their health activities, leading to long-term behavioral modifications towards a healthier dietary and life style, which ultimately leads to an improvement in their health outcomes (e.g., glucose values, hospital visits). This finding provides strong evidence of the underlying mechanism that drives the health outcome.

*Fourth*, in conjunction with patient self-management through the mHealth platform, we find heterogeneous effects between personalized and non-personalized messages. Interestingly, paired with all the health-management functions and resources provided by the mHealth platform, non-personalized SMS message interventions with general guidance about diabetes care demonstrate on average the highest effect on reducing patient glucose over time, 18.2% higher than personalized SMS message interventions with patient-specific medical guidance and 7.9% higher than no mobile message intervention at all.

Moreover, personalization is not as effective as non-personalization if we try to improve diabetes patients' engagement with the app usage or general life style (i.e., sleeping behavior or movement habits). This is likely because patients might perceive frequent personalized SMS messages as intrusive and annoying as mentioned in prior work albeit in a different context (Pop-Eleches et al. 2011). These findings are surprising and suggest personalized messaging may not always work in the context of mHealth, and the design of the mHealth platform is critical in achieving better patient health outcomes.

The major contributions of our study are as follows. *First*, to the best of our knowledge, our study is among the first research to examine the effectiveness and mechanism of the mHealth application platform on chronic-disease management. To disentangle the underlying mechanism that drives the observed health outcome, we investigated the detailed patient activities, such as daily walking steps, exercise time, sleeping pattern and food intake, documented through the app, together with the detailed app usage data. This step enables us to understand how and why mHealth technologies are able to lead to improved healthcare outcomes through patients' behavioral modifications.

*Second*, by partnering with a major mHealth platform as a real-world testbed, we design and conduct a randomized field experiment. This step enables us to identify and measure the impact of mHealth on patient health from a causal perspective, by eliminating the potential self-selection bias in mHealth technology adoption. Moreover, our randomized experiment was conducted on a relatively large scale (with eligible sample size n=1070), over three-month treatment period together with a follow-up after five months. This experimental design allows our findings to be more rigorous than most prior research that was conducted via smaller-scale pilot studies, over a shorter study period, or without follow-ups in the long run.

*Third*, this study also presents a unique opportunity to examine the potential economic impact of mHealth technologies on the efficiency of healthcare management.

*Fourth*, our research provides important insights on mHealth platform design through a better understanding of patient health behavior and interactions with the platform. Such knowledge can be highly valuable for healthcare mobile platform designers and policy makers to improve the design of smart and connected health infrastructures through sustained usage of the emerging technologies.

The rest of this paper is organized as follows. Section 2 discusses the related literature. Section 3 describes in detail how we design the randomized field experiments and how we partner with the real-world testbed to carry out the experiment on a large scale. Section 4 describes the experimental data. Section 5 discusses how we analyze the data as well as our final results. Section 6 discusses further analyses on patient activities and app usage to understand the underlying mechanism that drives the observed healthcare outcomes. Section 7 discusses additional robustness tests. Finally, Section 8 concludes with potential future directions.

## 2. Literature Review
### 2.1 Impact of Healthcare IT

Our work is related to prior literature on the impact of healthcare IT. Recently, with the development of healthcare IT technologies and digital platforms, researchers have looked into the digital transformation of healthcare (e.g., Agarwal et al. 2010, Bardhan et al. 2015, Liu et al. 2019). There has been growing interest in the consumer perspective of healthcare IT (e.g., Agarwal and Khuntia 2009, Gao et al 2010, Yan and Tan 2014, Bardhan et al. 2015, Liu et al. 2019,). Recent studies have examined the impact of healthcare IT on patient care outcomes (e.g., Anderson and Agarwal 2011, Bardhan and Thouin 2013). For example, Bardhan et al. (2015) focused on a chronic condition (congestive heart failure (CHF)) and examined health IT usage in relation to visits and readmissions. They found the adoption of health IT is associated with a reduction in the readmission risk of CHF patients. Interestingly, the evidence thus far for the impact of healthcare IT on patient care outcomes is equivocal, with prior research reporting positive, negative, and nonexistent effects (Agarwal et al. 2010, Bardhan et al. 2015). This, to a large extent, is due to the limitation in data deficiencies and limitations in the econometric estimation methods (Bardhan et al. 2015). These discrepant findings call for plausible explanations and present important opportunities for further work, especially from the patient care perspective.

Recently, studies have also focused on the social perspective and online healthcare platform design (e.g., Kane et al. 2009, Gao et al. 2012, Liu et al. 2019, Yan and Tan 2014, Yan 2020). For example, Liu et al. (2019) proposed an interdisciplinary lens that synthesizes deep learning methods to examine user engagement with encoded medical information in YouTube videos. They found videos with low medical information result in non-engagement; at the same time, videos with a greater amount of encoded medical information struggle to maintain sustained attention driven engagement. Yan and Tan (2014) investigate the role of social support from online healthcare community in patients' mental health. They found that patients benefit from learning from others. Yan (2020) further studies how online communities can better design social tools to facilitate communication and establish a variety of relationships between users.

In addition to the above, our paper is related to a stream of literature regarding the impact of healthcare IT on patient self-management of disease (particularly chronic disease).[6] For example, Lancaster et al. (2018) has reviewed the use and effects of recent electronic health (eHealth) tools

---

[6] For a review of recent studies on patient self-management, Allegrante et al. (2019) provided a systematic analysis of selected outcomes from randomized controlled trials of chronic disease self-management interventions contained in 10 Cochrane systematic reviews, which provided additional evidence to demonstrate that self-management can improve quality of life and reduce utilization across several conditions.

(including: linked to electronic medical record, personal health record, web-based surveys and drug list, web-based access to lab results, patient educational resources, patient-clinician messaging) for patient self-monitoring. They found consistent evidence that the use of eHealth tools can lead to improvement in patient symptoms. However, little evidence was found to support the effectiveness of eHealth tools at improving patient self-efficacy and self-management of chronic disease. And no evidence was found towards medication recommendations and reconciliation by clinicians, medication-use behavior, health service utilization, adverse effects, quality of life, or patient satisfaction (Lancaster et al. 2018).

Our study builds on this prior set of literature on the impact of healthcare IT. We distinguish our study by focusing specifically on the novel context of mHealth technology, and examine its impact from the patient care perspective, including patients' engagement with mHealth applications, patients' self-efficacy and self-management of chronic disease, patients' behavior modification and health outcome, and patients' healthcare costs. We also focus on understanding both the immediate impact (upon adoption and usage) and long-term impact (three to eight months after the adoption and usage).

## 2.2 Mobile Health (mHealth) and User Behavior

Our paper is also related to the recent work on mHealth and how it can change user behavior and adherence to medical treatment. Several recent studies have successfully piloted programs based on mobile SMS text messages, targeting patients with asthma, obesity, smoking, HIV/AIDS, and diabetes (e.g., Lester et al. 2010, Pop-Eleches et al. 2011). They have found an impact from mobile SMS messaging on user health behavior; however, the content, intensity, and delivery mode of the SMS messaging seem to have a significant influence on the effectiveness of the mHealth interventions (Free et al. 2013).

For example, Pop-Eleches et al. (2011) conducted a randomized trial using mobile SMS interventions in Kenya to test the effect of mobile SMS reminders on the adherence to HIV treatment.

They found simple weekly reminder messages (without any additional counselling) can significantly improve adherence. But surprisingly, more frequent daily messages do not improve patient adherence, because of potential habituation or intrusion. They also found adding more personal words, such as words of encouragement, in the longer text messages was not more effective than either a short reminder or no reminder.

A recent survey by Wang et al. (2017) systematically searched PubMed for mHealth-related studies on diabetes and obesity treatment and management published since 2000. They found existing studies in this area mainly focused on examining the impact from three major types of mHealth interventions: mobile phone text messaging, wearable or portable monitoring devices, and smartphone apps. They also noted that most existing studies included only small samples (<60 subjects per group) and short intervention periods (<3 months, no follow-up) and did not use rigorous data collection or analytic approaches. Although some studies suggest that mHealth interventions are effective and promising, most are pilot studies or have limitations in their study designs. There is an essential need for future studies that use larger study samples, longer intervention and follow-up periods to provide comprehensive and sustainable support for patients and health service providers.

Similar to our paper, a few recent studies also focused on examining the impact of diabetes smartphone app on patient health such as patient weight loss (Allen et al. 2013, Martin et al. 2015) or food intake (Nollen et al. 2014). Instead, our study focuses on more concrete healthcare-centric outcomes including blood glucose, hospital visits and medical expenses. Moreover, we also investigate various behavioral outcomes including patient activities and app usage to disentangle the underlying mechanism that drives the observed health outcomes. This is one unique feature of our study which distinguishes it from all existing studies. Another study related to ours is Rossi et al. (2009). The authors examined the impact of a mobile application "Diabetes Interactive Diary" on Type I diabetes patients but found the app was associated with a non-statistically significant reduction in blood glucose based on a study with only 41 patients. Lin et al. (2016) found a strong positive impact of the mobile-based visual diary and dietitian support on improving customer engagement. Using a unique dataset from a freemium mobile weight management application, Uetake and Yang (2017) have found the impact of short-term goal achievement varies across user segments.

Compared with these studies, our work distinguishes itself in its focus on mHealth app and chronic disease care (particularly diabetes), to examine the causal impact on patient behavior, medical expense, and health outcome. Our study is significantly different from all these papers in research context, goal, methodology, and study scale (sample size): (i) Our study focused mainly on Type II diabetes which, different from Type I diabetes, is directly tied to dietary or lifestyle self-management; (ii) Our goal is to understand the *causal* impact of mHealth app on diabetes patient health outcomes, as well as the underlying mechanism of how such technology can persuade patients to modify their behaviors to achieve these outcomes; (iii) our research method was based on randomized controlled trial, (iv) our study was conducted based on a much larger scale (sample size n=1070), which allow our study to be much more rigorous than many existing pilot studies.

**3. A Randomized mHealth Field Experiment**

To evaluate the effectiveness of the mHealth app on patients' behavior and health outcomes, one could collect secondary app user data and examine the user health behavior before and after the app adoption. However, the critical challenge for such an archival data analytical approach is the potential (strong) self-selection bias in the app user population. For example, users who care more about their health will be more likely to adopt the mHealth app, and will be more likely to change their behavior and life style in a healthier direction. This self-selection could lead to a statistically significant and positive correlation between the app adoption/usage and user health over time. However, this positive relationship might be endogenous, because of the unobserved user-level attributes that lead to the app adoption/usage in the first place.

Therefore, ideally, we would like the users to be randomly assigned to use the mHealth app—those who use the app and those who do not use the app will show no significant difference statistically. If so, the difference in their health behavior changes before and after the app adoption would be attributed solely to the impact of the app adoption/usage over time. Unfortunately, using only secondary data, we cannot easily identify such an impact from a causal perspective.

To ensure the random assignment of users, we propose to design and implement a randomized field experiment by partnering with a major mHealth company in Asia that provides the largest mHealth app platform in the nation that specializes in diabetes care. In this section, we will first introduce the background of this mHealth app platform. Then, we will discuss in detail how we design and implement our experiment.

**3.1 Mobile Health Platform Background**

Our research partner is a major mHealth firm in Asia. It provides the largest mHealth platform for chronic-disease management, specializing in diabetes care. To date, the mobile platform has 156,120 active users and 9,970 affiliated physicians who specialize in diabetes care across the nation. In addition to the external expert network, the platform also has a full-time internal expert team with more than 20 medical professionals including physicians, pharmacists, nurses, psychologists, and nutritionists. The platform integrates all the medical resources into a mobile app for patients.

This patient app provides diabetes patients with 24/7 services with four sets of core functions to facilitate patient self-management: (1) Behavior Tracking: patients can record and upload at any time their blood glucose, blood pressure, exercises, diet, weight, sleep, and so on. (2) Risk Assessment and Personalized Solutions: a cloud-based backend data analytic system will analyze individual patients' data and assesses the real-time health risk for each patient by taking into consideration 45 different types of medical conditions, including the stage and type of diabetes, whether the patient is pregnant, whether the patient has a complication, and so on. Based on the data analytic results, the app will recommend personalized self-management solutions for each patient regarding diet, exercise, life style, and potential medication. To ensure the validity of the recommendation, the internal medical team will view and discuss the data analytic results and personalized solutions regularly to improve the algorithm. (3) Q&A: the patients can contact the physicians in the internal and external expert networks for free consultation at any time regarding the medication, treatment, or self-management of their health. (4) Patient Community: the patients can participate in a digital community through the mobile app platform to discuss and communicate with each other.

For a better understanding of the patient app function, we provide screenshots of the major functions in Figure A1 in Online Appendix A. In particular, (1a) illustrates the overview of the user homepage after login. (1b) illustrates the page of recording a new blood glucose value. (1c) illustrates a set of user behavior tracking pages that visualize blood glucose, blood pressure, diet, and exercise. In addition, we also provide more screenshots for other related app functions in Figures A3 and A4 in Online Appendix A.

One critical challenge from the app platform designer's perspective is to examine how effective the app is in actually improving the patient health behavior and outcomes over time. To achieve this goal, we designed a large-scale randomized field experiment, which we discuss next.

**3.2 Experiment Design and Implementation**

We designed and implemented a nationwide large randomized field experiment by partnering with the firm. Our national campaign for the event received widespread attention from the society. To examine the impact of the mHealth platform under various situations, we designed five experimental conditions (2 Control groups + 3 Treatment groups) as follows:

- Control Group (C1): No treatment, behave as usual;
- Control Group (C2): Use the web (PC) version of the health app;
- Treatment Group (T1): Use the mHealth app;

- Treatment Group (T2): Use the mHealth app + Receive non-personalized SMS reminder messages with general knowledge about diabetes care twice a week; and
- Treatment Group (T3): Use the mHealth app + Receive personalized SMS reminder messages with patient-specific health advice from the internal expert team twice a week.

Control group C1 is the baseline. Control group C2 is a second baseline to examine the potential device effect that can lead to differences in the effectiveness of the diabetes self-management application. Treatment group T1 contains the normal mHealth app users who have access to all four sets of app functions. We designed treatment group T2 to test the potential synergetic effect when the mHealth app is paired with the mobile SMS messaging; research has shown the latter alone to be effective in improving patient treatment adherence and health outcomes (e.g., Lester et al. 2010). Finally, we designed treatment group T3 to further test the potential impact from the design of the SMS messaging, which were shown to have a significant influence on the effectiveness of the mHealth interventions (Pop-Eleches et al. 2011, Free et al. 2013). We provide an example of the two types of mobile SMS messages in Figure A5 in Online Appendix A. Notice that all three treatment groups have access to the same mHealth support including behavioral tracking, personalized risk assessment and solutions, Q & A, and online community. The SMS reminder messages sent in T2 or T3 do not contain any new information beyond what is shown in the app, but they simply serve as an additional "nudge" (either non-personalized or personalized).

We recruited participants for our experiment based on a voluntary basis through a combination of channels, including announcements through several national major news websites, social media and social networks via both web and mobile platforms, as well as offline recruiting through local hospitals and communities. Upon registration, each participant was randomly assigned to one of the five experimental groups. As compensation for their time and efforts, participants were automatically enrolled in a lottery upon completion of the experiment. The potential rewards from the lottery included Apple Watch, Fitbit smart bands, blood glucose meters, air purifiers, or gift cards with various values (from $5 to $750).

The initial round of participant recruitment started in May 2015. One practical challenge in medical trials is the potential delays in recruitment and the high rates of dropout, which might lead to uncertainty in the treatment effectiveness and might confound results (e.g., Watson and Torgerson 2006, Gupta et al. 2015). To ensure an effective sample size, we conducted the experiment by recruiting participants on a rolling "first-come-first-served" basis until the target sample size was met. Such an approach is common in medical trials (e.g., Gupta et al. 2015, Yeary et al. 2017, Myerson et al. 2018). Overall, the recruitment period spanned over seven months, from May 2015 to Dec 2015. To guarantee that long recruitment window would not introduce any confounding factors caused by time trend, we conducted an additional sub-sample analysis by selecting a subset of control and treatment groups who were recruited into our experiment during the same month. We provide more details on this analysis in Section 7 for robustness checks.

The treatment period of the experiment lasted for three months (90 days) starting from the day of registration. Based on the random assignment to the experimental group, each participant received the corresponding treatment according to the experimental design during the treatment period. In addition, to collect patient-level demographics and medical history, as well as to evaluate the potential economic impact of the mHealth platform on patients' medical costs and hospital visits, we conducted additional surveys through telephone interviews before and after the treatment period. In particular, we interviewed each participant three times—first at the beginning of the experiment (during registration), and second at the end of the 3-month treatment period, and then another five months after that.

Therefore, for each participant, the total experimental period lasted for 8 months (i.e., pre-treatment survey + 3-month treatment period + post-treatment survey + 5-month post-treatment period + another post-treatment survey). Overall, the entire study for all our participants spanned 15 months from May 2015 to July 2016. The last batch of participants was recruited in December 2015. They completed the experiment and surveys by the end of July 2016.

During the three telephone interviews for the pre- and post-treatment surveys, we asked the participants about their demographics, medication and medical history, most recent blood glucose and glycated hemoglobin levels, frequency of hospital visits, medical costs, and so on.[7] Informed consent was obtained at each phase of the study that required data collection. In the next section, we will discuss in more detail the exact survey variables we collected.

Note that to eliminate potential confounding factors, during the experimental period we ensured the following facts: (1) no participant had previously adopted the mHealth app prior to the registration to our experiment; (2) participants who were assigned to the two control groups did not happen to adopt the mHealth app during the experiment on their own. (We validated these first two facts by crosschecking the phone numbers between the participants and the mHealth app adopters in the company database, and also through the post-treatment survey to exclude those who were not supposed to be adopters of the app prior or during the experiment.) (3) participants did not adopt other similar apps during the experiment. (We validated this fact through the post-treatment survey to exclude the potential impact from other similar apps.) (4) participants had no other major medical conditions prior to the experiment. Besides, we focused on Type II diabetes which, different from Type I diabetes or Gestational diabetes, is directly tied to dietary or lifestyle self-management. Finally, to avoid potential bias due to misalignment with participants' prior expectation, we followed prior social and behavioral research methods (Hoyle et al. 2001) and ensured that the recruitment announcement only revealed the general purpose of the experiment (i.e., to help improve diabetes care), whereas it did not reveal the exact details of the experiment (i.e., to study the impact of adoption of mHealth app on diabetes patient behavior).

---

[7] We provide the details about the pre- and post-treatment surveys in Online Appendix B.

**4. Data**

In this section, we will describe our data from both the experiment and the pre- and post-treatment surveys. We first illustrate our data sampling procedure during the recruitment and randomization processes. To validate our samples, we conducted the randomization check and briefly discuss it.

**4.1 Randomization and Sampling**

Our recruitment process led to the enrollment of 1,770 patients. To ensure minimum confounding factors, we excluded 427 (24.1%) patients from our sample who did not have diabetes (e.g., people whose blood glucose value was reaching the upper bound of the normal range but were not classified as diabetic yet), or had other major chronic disease(s) at the same time (e.g., kidney disease, heart disease, arthritis, HIV/AIDS), or were already users of the app. These exclusions led to a sample of 1,343 patients whom we randomly assigned into one of the five experimental groups. During the three-month treatment period, 273 (15.4%) patients dropped out. Hence, our final eligible sample for analysis contains 1,070 patients, 60.5% of the original enrolled sample. We illustrate the flow of the randomization and sampling procedure in Figure C1 in Online Appendix C.

Note that high patient dropout rate is a common challenge in medical trials (e.g., Gupta et al. 2015). To alleviate any additional concern towards this issue, we compared the distributions of participants' demographic and baseline health-related characteristics between the dropout samples and the eligible samples. We did not find statistically significant difference between the two. We also compared the distributions of participants' demographic and baseline health-related characteristics among all the dropout samples across the five experimental groups. We did not find statistically significant difference across the control and treatment groups regarding dropout samples. In Section 7, we provide more detailed results on these tests. Therefore, while we acknowledge this fact as one potential data limitation in our study, we are more confident that it is not a serious concern in affecting our results.

**4.2 Data Description**

Our main experimental data contain a combination of three data sets:

(1) Panel data of individual health and behavior characteristics recorded through the mobile (or web-based) health application during the three-month treatment period. This information includes diabetes-related health activities such as glucose value, glucose type (e.g., pre-/post-breakfast, pre-/post-lunch, pre-/post-dinner, before sleep), and uploading time/date. Notice that for the control group (C1) that did not use the mobile or web-based health application, we asked the participants to upload their glucose values at least twice: at the beginning and end of the three-month treatment period through a web portal. We provide the screenshot of this web portal in Figure A2 in Online Appendix A.

(2) Panel data of individual activities and app usage logs. This information includes walking steps, exercise time and calories burned, food intake and estimated calories, sleeping time (starting and ending time, and length), app opening time and frequency, frequency of documenting activity logs, loyalty rewards, shopping activities (product purchased, price, order time), in-app Q&A with medical experts (query time, answer time). For some of the activities such as walking steps, the app can automatically log them through the build-in sensors of the smartphone.[8] For other activities like exercise, sleep or food intake, they require the patients to document them in the app. Note that the patients only need to document (select from a pre-compiled list) the type of exercise/food and corresponding time/amount, the app can then automatically calculate the estimated calories burn/intake. For the purpose of understanding patient app usage behavior, we consider the frequency of documenting activity logs as the times only when patients document exercise, sleep and food activities in the app (instead of the automatic activity logs generated by the app).

---

[8] For patients in control group C2 (who were asked to document all the activities through the web-based application on the PC), it was difficult to record the walking steps by the patients themselves. Hence, we suggested them use information from any other movement tracker's (such as iPhone's inherent Health app or Xiaomi's Mi Fit app) if they had any. We found two patients did not have such information. We excluded them later from the corresponding analysis on #steps.

(3) Survey data of individual demographics, health, and behavior characteristics from the pre- and post-treatment surveys. This information contains individual age group, gender, marital status, income level, diabetes type (i.e., type 1, type 2, gestational), diabetes age (time since diabetes was first diagnosed), frequency of glucose monitoring, whether the patient has any complications, the most recent blood glucose value and type, glycated hemoglobin for the most recent three months, average time for exercise and sleep per day during the most recent three months, average calories per meal during the most recent three months, whether the patient is a smoker or drinker, whether the patient is pregnant or not, current and past medication, medical history (e.g., blood pressure, blood fat, family history), frequency of hospital visits per year, frequency of hospital visits during the last three months, and medical costs during the last three months. The survey data also contain information on individual app-related activities including registration time/date, frequency of app daily usage, and satisfaction rate. For details on these variables, we provide the summary statistics in Table 1.

**4.3 Randomization Check**

To validate the randomization procedure, we conducted a randomization check. We provide the details about the randomization check in Table 2. Across the five experimental groups, we compared the distributions of the patient demographics and baseline health condition characteristics. We found the distributions are similar across groups. Furthermore, to better control for the potential variation in the patient-level characteristics, we tested several different models by including all or different subsets of these variables in our analyses as control variables. We found our results stay highly consistent. We will discuss more details in the next section.

## 5. Analysis and Findings

In this section, we discuss how we analyzed the experimental data to examine the impact of the mHealth platform on patient health behavior and outcomes. Note we have both the panel data on patient health and behavior characteristics during the three-month treatment period, and the cross-sectional survey data before treatment (upon registration) and five months after treatment. We first conduct a group-level analysis using the survey data to compare the difference in patient health and behavior before and after the treatment. Then, we use the panel data to conduct the analysis of the treatment effect at the individual level.

### 5.1 Group-Level Analysis

First, we conduct a group-level analysis using the survey data to compare the difference in patient health and behavior before and after the treatment. Note the total time period between the two surveys is eight months: a three-month treatment period plus a five-month post-treatment period. By doing so, we aimed to capture the potential long-term effect of the treatment. In particular, across the five groups, we compare the differences in the blood glucose and glycated hemoglobin levels, the number of hospital visits during the most recent three months, and the total medical spending related to diabetes during the most recent three months. We provide the details in Table 3. The values across groups are statistically different at the $p<0.05$ level based on the one-way ANOVA test.

The first thing we notice is that in the baseline control group (C1), the four variables stayed relatively stable before and after the treatment, whereas all other groups that used the health application (whether mobile- or web-based) showed a significant reduction in patient glucose and hemoglobin values, as well as a reduction in hospital visits and medical spending. This finding is promising. It indicates the health platform for diabetes self-management indeed has a significant effect on improving patient health outcomes as well as reducing costs.

Second, compared to the second baseline group (C2) with web-based health intervention, the three treatment groups with mHealth interventions (T1, T2, T3) experienced a statistically significantly higher impact on patient health and costs. For example, under the same functional setting of the health application, we observe a 21.6% increase in the mobile-based platform's (T1) impact on reducing patients' glucose, compared with the web-based platform's (C2) impact. This result is consistent with previous findings indicating a significant mobile device effect (e.g., Xu et al. 2016, Wang et al. 2016, Jung et al. 2019). Such an effect can become salient in personal health management through faster and more flexible user response to real-time information and mobile-enhanced user self-efficacy (e.g., Lin et al. 2016).

Third, we notice that among the three mobile treatment groups, T2, when we paired the mHealth app with simple non-personalized SMS reminder messages about general guidance on diabetes care, demonstrates the strongest treatment impact on reducing blood glucose levels over time, 18.2% higher than personalized SMS message interventions with patient-specific medical guidance and 7.9% higher than no mobile message intervention at all. We also see a consistent trend in the Hemoglobin value. Interestingly, T3, when we paired the mHealth app with personalized SMS messages about patient-specific medical advice, does not perform better than non-personalized messages in helping patients improve their health outcome. This finding is surprising but highly consistent with prior research that the design of the SMS messaging has a significant influence on the effectiveness of the mHealth interventions (Free et al. 2013), and that more personal and encouraging words in longer text messages were not more effective than either a short reminder or no reminder, because of potential habituation or perceived intrusion (Pop-Eleches et al. 2011), and that personalization might lead to potential privacy concerns and information overload for consumers (e.g., Aral and Walker 2011, Goldfarb and Tucker 2011, Ghose et al. 2014, Ghose 2017).

Finally, and interestingly, when looking into the patient hospital visits and medical spending, we find T3 demonstrates the highest impact in reducing the two. T3 is 62.5% and 168.4% more effective compared with T2, the next best treatment, in reducing hospital visits and medical spending, respectively. This result suggests the potential of the mHealth app combined with personalized SMS messaging to reduce the medical and operational costs for chronic disease patients and healthcare providers. An intriguing implication of this finding is that, although personalized messaging is not more effective in affecting patient health outcome than non-personalized messaging, it can actually facilitate a personal connection between patients and physicians, which in turn leads to increased patient trust in the mHealth platform and higher willingness to adopt telehealth (e.g., through Q&As and online communications with physicians), hence reducing patients' need (or urge) to visit hospitals in person.[9]

Note that all the analyses in this subsection are based on the cross-sectional survey data and are conducted at the group (mean) level. The impacts here should be interpreted as the group-level mean treatment effect. To further account for the potential heterogeneity within the group, we conducted individual-level analysis using the panel data, which we will discuss next.

**5.2 Individual-Level Diff-in-Diff Analysis**

To better control for the potential individual heterogeneity and explain the potential discrepancy in the observed outcome, we conduct individual-level analysis using the panel data of individual health and behavior characteristics we collected during the three-month treatment period. Because our recruitment is conducted on a rolling basis, we consider the time indicator in our context as the time elapsed since the patient started the experiment. Particularly, in our analysis, it is defined as the unique sequence index of each patient's uploaded glucose value.

---

[9] To further verify this finding, we conducted a follow-up analysis to examine the potential usage of telehealth across different experimental groups (C2, T1, T2, T3). In particular, we looked into the frequency of in-app patient-physician communications. We found patients in T3 group indeed demonstrated the highest frequency of patient-physician communications through the platform, with an average of 4.35 times per person during the 3-month experimental period. In contrast, patients in T2 and T1 groups had an average of 2.57 and 1.44 patient-physician communications, respectively. We also found a similar mobile device effect in affecting patients' likelihood to adopt telehealth. Patients in C2 showed the lowest frequency with only 0.78 patient-physician communications. A pair-wise t-test demonstrates these group-level differences are statistically significant.

To account for the patient-level baseline time trend [10], we apply a diff-in-diff method to model individual-level glucose change over time. In particular, the first-level difference is the within-group glucose change over time (i.e., group-specific time trend), and the second-level difference is the discrepancy in this time trend across groups. Put more formally, we model the glucose value $Glucose_{it}$ for patient $i$ at time $t$ as follows:

$$Glucose_{it} = \beta_0 + \beta_1 Treatment_i + \beta_2 Time_t + \beta_3 Treatment_i \times Time_t + X_i \beta_4 + C_{it} \beta_5 + \varepsilon_{it}, \quad [1]$$

where $Treatment_i$ represents the indicators of the five experimental groups. $Time_t$ represents the time indicator of how many days since the start of the treatment period when the corresponding glucose value was uploaded ($1 \leq Time_t \leq 90$).[11]

$X_i$ is a vector of control variables for patient-specific time-invariant characteristics including age group, gender, income level, marital status, diabetes type, diabetes age, frequency of glucose monitoring, whether the patient has any complications, most recent glucose and glycated hemoglobin levels prior to the experiment, average time for exercise and sleep per day and average calories per meal prior to the experiment, whether the patient is a smoker or drinker, whether the patient is pregnant, whether the patient has any other health concerns, such as high blood pressure or cholesterol, whether the patient is currently on any medications, and whether any patient-physician interaction occurred during the three-month treatment period.

$C_{it}$ is a vector of control variables for patient-specific time-varying characteristics including the time of day (morning, afternoon, evening), day of the week (Monday ~ Sunday), and month indicators of the corresponding glucose uploading activity, uploaded glucose type, daily exercise from the patient (total steps), as well as the patient's frequency of daily app usage (including all types of activities). $\varepsilon_{it}$ is a stochastic error to capture any randomness in patient behavior. The unobserved error term is assumed to be orthogonal to other independent variables and has a mean zero. In the estimation, we cluster the $\varepsilon_{it}$ at the experimental group level to account for potential within-group relationships.[12]

---

[10] We first examined the overall time trends in each experimental group regarding the blood glucose change over time at the individual patient level. We plot the glucose value over time for each group in Figure D1 in Online Appendix D.
[11] We use logarithm of $Time_t$ in our analyses to normalize the scale.

We have tested different models (Models I ~ IV) with different combination of the set of control variables. We provide our estimation results from these models in Table 4. In the estimation, the primary coefficient of interest is $\beta_3$, which is a vector that contains coefficients for the four interaction effects ($Treatment_i \times Time_t$). Note the control group indicator C1 is dropped due to collinearity (i.e., the interaction effect between C1 and $Time_t$ will be captured as the baseline effect, $\beta_2$, the coefficient of $Time_t$).

All four models demonstrate similar estimation results and provide evidence consistent with our previous group-level analyses. First, we notice that all four groups (C2, T1, T2, T3) experience a significant reduction in patient glucose values. This finding indicates the diabetes self-management platform (whether mobile- or web-based) is effective compared with the baseline control group (C1) that did not use the platform.

Second, comparing T1 with C2, we notice a significant device effect: the mobile-based platform is more effective than the web-based platform in reducing glucose levels over time. In particular, we find that when the time since the patient adoption of the platform doubles, the glucose value on average drops 0.24 mmol/L for patients in C2 and 0.28 mmol/L for patients in T1.[13]

Third, when comparing the three treatment groups (T1, T2, T3), we see an interesting trend: T2 (the mHealth app with non-personalized mobile SMS reminder messages) is overall most effective in helping patients reduce their glucose over time, whereas T3 (mHealth app with personalized mobile SMS messages) is less effective. In particular, our results show that when the time since mHealth adoption doubles, the glucose value on average drops 0.28 mmol/L, 0.32 mmol/L and 0.26 mmol/L, for patients in T1, T2 and T3, respectively. In essence, this indicates that T2 (non-personalized) exhibit a 23.1% higher effectiveness in decreasing patient glucose than T3 (personalized). [14]

---

[12] We also estimated the model without the clustered errors, and found the results are qualitatively consistent.
[13] We computed these marginal effects based on standard interpretation of the log-linear regression coefficients.

This observation is consistent with our findings from the group-level analyses as well as the prior literature (e.g., Harle et al. 2012), indicating the design of the mobile SMS messaging plays an important role in the effectiveness of the mHealth interventions on patient health outcomes (e.g., Free et al. 2013, Pop-Eleches et al. 2011). Carefully designing the content, format, intensity, and delivery mode of the SMS messaging is critical.

Finally, when looking at the baseline coefficients in Table 4, we see the four baseline coefficients ($\beta_1$) for the treatment groups ($C_2, T_1, T_2, T_3$) are not statistically significant. This finding further validates our random group assignment indicating the initial glucose values do not vary significantly across groups. Moreover, when looking at the baseline coefficient for $Time_t$, we find $\beta_2$ is statistically significant and positive for all groups. This finding indicates the baseline time trend of patient glucose for control group (C1) without any intervention is increasing over time. This result delivers an important message. It indicates the potential risk and challenge in diabetes care over time, and suggests the importance of empowering patients to improve their self-management for diabetes through smart and digital health platforms.

**5.3 Patient-Level Fixed Effect**

In the previous section, we considered a large number of patient-level characteristics in the individual-level analysis to control for individual-level heterogeneity. To further account for any other potential unobserved individual characteristics, we conduct the diff-in-diff analysis with patient-level fixed effects as follows:

$$Glucose_{it} = \beta_0 + \beta_1 Time_t + \beta_2 Treatment_i \times Time_t + \mu_i + \boldsymbol{C_{it}\beta_3} + \varepsilon_{it}, \qquad [2]$$

where $\mu_i$ captures the patient-level fixed effect. Note that in this model, we drop the treatment group indicator $Treatment_i$ and the patient-specific time-invariant characteristics $\boldsymbol{X_i}$ from the model because of collinearity with the patient fixed effect. The primary coefficient of interest is $\boldsymbol{\beta_2}$, the interaction between the treatment group indicator and time. We estimate the model with the patient-specific time-variant characteristics, $\boldsymbol{C_{it}}$ (Model V), and without, $\boldsymbol{C_{it}}$ (Model VI). The corresponding estimation results are shown in Table 5.

---

[14] The difference between T2 and T3 is computed directly by (0.32-0.26)/0.26 = 23.1%.

Overall, our findings from the patient-level fixed-effects model demonstrate high consistency with our previous analysis using the treatment-group-level fixed effect (i.e., equation [1]). We find the adoption of the mobile-based platform (T1, T2, T3) can statistically significantly improve the health outcome of diabetes patients in reducing their blood glucose values over time, even after controlling for the individual-level fixed effects.

Moreover, we also see a consistent trend: in conjunction with the mHealth app platform, non-personalized mobile messages with general guidance for diabetes care have a higher impact on patient health improvement than personalized mobile messages. These additional empirical analyses provide us with robust evidence in our results.

## 6. Further Analyses on Patient Behavioral Modifications

To disentangle the underlying mechanism that drives the observed health outcome, we further investigated the detailed patient behavioral activities, such as walking steps, exercise time, sleeping pattern and food intake, documented through the app, together with the detailed app usage data. This step enables us to understand how patients actually use the mHealth app and what kinds of behavioral modifications occur in response to the app usage. Note that looking into the detailed patient activities and app usage logs to study how exactly mHealth technologies can lead to patients' behavioral modifications over time to achieve better healthcare outcomes is a unique feature of our work, which distinguishes it from all the existing work in this area.

**6.1 Analyses on Patient Activity and App Usage**

We conducted empirical analyses to study the impact of mHealth treatments on each of these patient activity and app usage outcome variables, using a Diff-in-Diff model with patient-level fixed effect. We provided the detailed estimation results in Tables 6a and 6b (Patient Activities) and Table 7 (App Usage). Note that because control group C1 did not have access to the health application (web or mobile), we did not have any individual-level activities or app usage data from these patients. In all the analyses below, control group C2 (who had access to the PC-based application) was used as the baseline for comparison.

More specifically, our main findings are the following. First, when looking into the patient activities as outcome variables (Tables 6a and 6b), we found that compared to patients from the PC group (C2), patients from the three mHealth treatment groups (T1, T2, T3) did significantly higher level of daily exercise, consumed healthier food with lower daily calories intake, walked more steps and slept for longer time a day. For example, based on Table 6b[15], when patients double their time of mHealth app usage, we observe an average of 16.8% decrease in food calorie consumption, 6% increase in daily exercise time, and 14% increase in daily sleeping time, for treatment group T1. We observe a similar trend for T2 and T3, with an average of 17.2% (T2) and 17.1% (T3) decrease in food calorie consumption, 6% (T2) and 4.3% (T3) increase in daily exercise time, and 18.2% (T2) and 11.1% (T3) increase in daily sleeping time.

---

[15] Note that because of the presence of control variables, Table 6b will give us more conservative estimates compared to Table 6a.

These findings indicate that patients indeed have made significant behavioral modifications towards a healthier dietary and life style after adopting and using the mHealth application. As seen from our results, patients in the mHealth treatment groups became more autonomously self-regulated with their health behavior. Such increasing intrinsic motivation helped them become more engaged, persistent and stable in their behavior, leading to an improvement in their health outcomes (e.g., glucose values, hospital visits).

Interestingly, we found the three mHealth treatment groups performed relatively similarly in daily food calories intake from breakfast, lunch and dinner. However, we noticed a significant drop in the performance from T3, the patient group provided with additional personalized SMS reminder messages, with regard to daily walking, exercise and sleeping patterns. For example, based on Table 6b, when combining mHealth app with personalized reminder messages (T3), it leads to a 15% decrease in the number of daily walking steps and a 27.8% decrease in total exercise time compared to using mHealth app alone (T1), and it leads to a 41.2% decrease in the daily steps and a 28.4% decrease in total exercise time compared to combining mHealth app with non-personalized reminder messages (T2).

Meanwhile, providing additional personalized reminder messages (T3) also leads to a 20.5% and a 39.1% decrease in daily sleeping length, compared to using mHealth app alone (T1) and providing additional non-personalized reminder messages (T2) respectively. Furthermore, when looking into the frequency of late night sleep – when patients went to sleep later than 11pm – we noticed that providing personalized reminder messages in conjunction with the mHealth app can lead to more frequent late night sleep by the patients.

These findings suggest that highly personalized messages may not always work well in trying to persuade patients' behavioral modifications. As shown in our results, personalization is not as effective as non-personalization if we try to improve diabetes patients' general life style (i.e., sleeping behavior or movement habits).

This is likely because patients might perceive frequent personalized SMS messages as intrusive and annoying (Pop-Eleches et al. 2011). More importantly, frequent personalized messages might cause patients to feel pressured or coerced by intrapsychic or interpersonal forces, which can significantly demotivate patient behavior from being autonomously self-regulated (Deci and Ryan 1985, 2000).

Second, when looking into the app usage as outcome variables (Table 7), we found that overall patients from the three mHealth treatment groups (T1, T2, T3) demonstrated a higher level of usage activities compared to the PC group (C2) – opening app and documenting their daily health activities more frequently, more frequent in-app communications with medical experts, and higher loyalty rewards. This finding indicates a strong positive impact of mobile platform on patient engagement with the healthcare technologies. Patients are more likely to engage with the health self-management functions provided in a more flexible setting (i.e., on mobile devices). The accessibility, convenience, and ubiquity inherent to mobile devices help patients easily upload information on a regular basis and follow the guidance that would eventually lead to improved health conditions.

Among the three mHealth treatment groups, the effect appeared to be the strongest when combining the mHealth app with non-personalized reminder messages (T2), followed by the case when using mHealth app alone (T1). Again, we found that providing additional personalized reminder messages can attenuate the mHealth treatment effect and lead to lower app usage by the patients – lower daily frequency of opening app and documenting health activities, lower frequency of communicating with medical experts, and lower loyalty rewards.

This is likely due to patient perceived intrusion, annoyingness and privacy concern (Pop-Eleches et al. 2011). Moreover, frequent personalized messages might cause the patients to feel increased control and judgment. Lack of choiceful and volitional feeling can lead to loss of autonomy and self-motivation (Deci and Ryan 1985, 2000). And correspondingly, it can lead to lower engagement in app usage.

**6.2 Mediation Effect of Patient Behavioral Change**

In addition to the above analyses, to further test the mediation effect of patient behavioral change on the health outcome, we conducted two additional mediation analyses using (1) a simultaneous equation model, and (2) a directed acyclic graph (DAG) in the form of a parametric structural equation model (SEM).[16]

First, we applied a simultaneous equation model to analyze the health outcome and the patient activities simultaneously. More specifically, we model the *glucose change* (i.e., post experiment – pre experiment) for each patient as a function of individual *behavioral activities* (i.e., exercises, food intake), demographics, and other control variables ($X_i, C_{it}$); in the meantime, we model the individual behavioral activities as a function of mHealth app treatment, while controlling for demographics and other factors ($X_i, C_{it}$).

$$GlucoseChange_{it} = \gamma_0 + \gamma_1 BehavioralActivities_{it} + X_i \gamma_2 + + C_{it} \gamma_3 + \omega_{it}, \quad [3]$$

$$BehavioralActivities_{it} = \alpha_0 + \alpha_1 Treatment_i + X_i \alpha_2 + C_{it} \beta \alpha_3 + \mu_{it}. \quad [4]$$

We provide the results in Table 11 and Table 12. As we can see from this analysis, the effects of mHealth adoption (T1, T2, T3) demonstrate a statistically significant impact on increasing patients' exercises and decreasing patients' food calories intake, which in turn, leads to a lower blood glucose over time.

Second, we built a directed acyclic graph (DAG) in the form of a parametric structural equation model (SEM) to test the causal path of the mHealth impact on patient health outcome through the behavioral modification. In particular, we empirically test whether there is a statistically significant impact of mHealth adoption through the mediation effect of individual behavioral activities (i.e., exercises, food intake). We provide the estimation results in Figure 1 and Figure 2.

---

[16] Note that because control group C1 did not have access to the health application, we did not observe any individual-level behavioral activities from these patients. In the mediation analyses, control group C2 (who had access to the PC-based application) was used as the baseline for comparison.

In Figure 1, C1 is dropped out of the model (hence no coefficient estimated associated with the arrow) because no behavioral activities were observed for this group. C2 is dropped out of the model because it is used as the baseline. The effects of mHealth adoption (T1, T2, T3) are highly consistent with our main analyses, demonstrating a statistically significant and positive impact on patients' exercise activities (with the estimated coefficients 0.81, 0.88, 0.26, respectively), which in turn, leads to a lower blood glucose over time (with the estimated coefficient -0.14). We also found a consistent trend where the combination of non-personalized SMS (T2) was the most effective, whereas the personalized SMS (T3) was the least effective.

We also conducted similar empirical test for the causal mediation effect of "Food Intake" and found consistent trend. As we see in Figure 2, the effects of mHealth adoption (T1, T2, T3) demonstrate a statistically significant and negative impact on patients' food calories intake (with the estimated coefficients -0.4, -0.26, -0.21, respectively). In the meantime, lower food calories intake leads to a lower blood glucose over time (with the estimated coefficient 0.19).

Overall, we found that the two additional mediation analyses using the simultaneous equation model and the directed acyclic graph (DAG) have demonstrated highly consistent evidence with our main results. They further support the causal impact of mHealth adoption on the health outcome, through the mediation effect of patient behavioral change.

### 6.3 Additional Follow-up Survey and Interview

To further verify our findings, we conducted an additional round of follow-up survey and interview. We provide the details in Online Appendix E. We asked the participants three major questions: (1) What is your favorite function of the blood glucose management mobile app? (2) How did these functions help improve your health? (3) For participants in T3, what's your feedback towards the personalized text messages about medical guidance based on your personal exercise, diet and health status?

Based on the survey user responses, the most useful app function liked by the users is "*Learning about health knowledge* (68%)," followed by "*Health real-time tracking: blood glucose, exercise, drug, diet* (67%)," "*Personalized diabetes risk assessment* (61%)," "*Doctor consultation* (53%)," and "*Social network support* (48%)."

When being asked how these functions helped improve their health, a large majority of the users mentioned that the app provided them a way to better monitor and "quantify" their life and health in real time, hence they were able to better manage food intake and exercise. For example, *"The combination of my blood glucose level and exercise diet allows me to understand the relationship between them clearly, which motivates me to exercise more and eat healthier food"; "Self-tracking of health status provides a quantitative basis in real time, and thus improves my health level."*

Regarding the personalized text message about medical guidance, users raised three major concerns: (1) *Interruption and annoyingness* (58%), (2) *Avoidance towards negative information* (54%), and (3) *Privacy* (53%).

In addition to the survey responses, we have also conducted in-depth phone interviews with a randomly selected group of 7 experimental users from T3 treatment group. The main purpose of the interview was to further verify the survey responses, and meanwhile with a focus on why the personalized text messages did not work well.

We found the responses from the interview were highly consistent with those from the survey. When the participants were asked what functions they liked the most, all of the 7 interview participants indicated that the real-time health tracking function provided them a way of better monitoring and managing their health. They (and their family members) have also gained professional health knowledge through using the app.

When the participants were asked whether they liked the personalized medical guidance via text messages and why, a majority of them (6 out of 7) indicated that they found these personalized text messages "too frequent," "annoying" and violating "privacy."

Interestingly, during our interview one of the participants explicitly mentioned his/her preference of a less personalized text message to avoid "being judged all the time by someone." This is highly consistent with our previous finding that frequent personalized messages might cause the patients to feel increased control and judgment. They can in tern lead to a significant decrease in patient intrinsic motivation of disease self-management and a lower health outcome.

**6.4 Summary of Findings and Managerial Implications**

Overall, our further analyses on patient glucose values, behavioral activities and app usage, together with the additional survey and interview, demonstrate highly consistent evidence that mobile health app platforms have a statistically significant impact on empowering patients with diabetes self-management, reducing patients' glucose values, improving their life style and health outcomes over time. Our results also provide strong evidence of the underlying behavioral mechanism that drives the observed health outcome.

More specifically, first, we find the adoption and usage of the mHealth platform has a significant impact on improving diabetes patient health outcomes as well as reducing medical costs. Second, between web-based and mobile-based platforms, we find a strong device effect: the mobile interventions led to a statistically significantly higher impact than the web-based intervention. Third, the mHealth platform also demonstrates a significantly stronger impact on patients' dietary and life style improvement as well as engagement with app usage than does the web-based platform. This finding suggests that patients in the mHealth treatment groups indeed became more engaged, motivated, and autonomously self-regulated with their health behavior over time. Such increased intrinsic motivation can in turn lead to an improvement in their health outcomes (e.g., glucose values, hospital visits). This insight is critical. It provides strong evidence of the underlying mechanism that drives the observed health outcome, demonstrating the potential of mHealth in empowering diabetes patients for efficient health management.

Furthermore, in conjunction with patient self-management through the mHealth platform, we find heterogeneous effects between personalized and non-personalized messages. Interestingly, paired with all the health-management functions and resources provided by the mHealth platform, non-personalized SMS messages demonstrate on average the highest effect on reducing patient glucose over time. In contrast, personalization is not as effective as non-personalization if we try to improve diabetes patients' engagement with the app usage or general life style (i.e., sleeping behavior or movement habits). This is likely due to patient perceived intrusion, annoyingness and privacy concern (Pop-Eleches et al. 2011). Furthermore, frequent personalized messages might cause the patients to feel increased control and judgment. They might cause patients to feel pressured or coerced by intrapsychic or interpersonal forces. We have seen such evidence in both our experimental analyses and our additional follow-up survey and interview (Online Appendix E).

Our finding is also highly consistent with prior research on Self-Determination Theory (SDT) and Cognitive Evaluation Theory (CET). Prior theoretical literature has demonstrated that lack of choiceful and volitional feeling can lead to loss of autonomy and self-motivation (Deci and Ryan 1985, 2000). It in turn can lead to a significant decrease in patient intrinsic motivation of self-management—demotivate patient behavior from being autonomously self-regulated in health. Such loss of autonomy can lead to lower patient engagement in the health self-management process (e.g., lower mHealth app usage, lower patient-physician engagement, lower compliance to medication and treatment). Also, prior research showed that pressured evaluations and imposed goals diminish intrinsic motivation because they conduce toward an external perceived locus of causality. In contrast, choice, acknowledgment of feelings, opportunities for self-direction, and positive social-contextual events (e.g., feedback, communications, rewards) were found to enhance intrinsic motivation because they allow people a greater feeling of autonomy and competence (Deci & Ryan, 1985).

In sum, these findings are surprising and suggest frequent personalized mobile messaging may undermine the effectiveness of mHealth technology. The design of the mHealth platform, and health IT in general, is critical in achieving better patient engagement and health outcomes.

Our findings have several important implications. From the healthcare provider's perspective, our study illustrates the importance of mHealth technology in facilitating diabetes patient self-management to improve well-being and health outcomes through behavioral modifications over time. Importantly, mHealth technology has shown great potential to improve patients' compliance - following diets and executing life style changes that coincide with healthcare providers' recommendations for health and medical advice.

From the mHealth platform designer's perspective, our study suggests the design of the mHealth platform is critical in achieving better patient engagement, empowerment, and health outcomes. Instead of personalized messaging, mHealth applications should be paired with non-personalized messaging with general knowledge about disease management for patient education. Our research also provides important design guidance for supporting communication and shared decision making via reminders, notifications and informed guidance, and improving care delivery operations to increase satisfaction and quality of care.

Finally, from the policy maker's perspective, our findings demonstrate the potential of mHealth technologies in improving healthcare delivery to significantly impact outcomes, quality and costs. Our study also significantly improves the understanding of issues that interfere with patients' sustained engagement with mHealth apps and population adherence to treatment and wellness regimens. It provides key insights in the underlying mechanisms that drive individual and population health behaviors and life style changes through mHealth app, and moreover, the critical policy implications regarding the adoption and sustained usage of mHealth technologies.

## 7. Robustness Analyses

We conducted several robustness analyses to check the validity of our experimental design and data quality. We discuss them in this section.

### 7.1 Long Recruitment Window

The rolling recruitment process in our study lasted for seven months. To guarantee that such long window would not introduce any confounding factors caused by time trend, first we have considered a time fixed effect in the individual-level Diff-in-Diff analysis (Time_t, coded as the time sequence index of patient glucose upload time) to control for any individual-level time trend. Moreover, in the analysis we have also controlled for the glucose type (before/after breakfast/lunch/dinner/sleep), the actual time, day and month indicators for the glucose upload time. This aims to control for any potential common time-of-day or seasonality effects for the entire population.

To further alleviate the concern, we have conducted an additional subsample analysis by selecting a subset of control and treatment groups who were recruited into our experiment during the same month. In particular, we focused on only those patients who were recruited in May 2015 (i.e., we chose the first month of the recruitment period to also minimize any potential risk of sample contamination). This led to a subsample of 285 patients: C1(n=49), C2(n=63), T1(n=57), T2(n=64), T3(n=52). We then conducted Diff-in-Diff analysis to compare the group means in the glucose change based on this subsample. Overall, our findings remain highly consistent based on the subsample analysis. The detailed results are provided in Table 8.

### 7.2 Sample Dropout

Indeed, high patient dropout rate is a common challenge in medical trials (e.g., Gupta et al. 2015). To alleviate any additional concern towards this issue, we conducted two levels of analyses:

First, we compared the distributions of participants' demographic and baseline health-related characteristics between the dropout samples and the eligible samples. Based on a Welch's t-test, we could not reject the null hypothesis that there is no statistically significant difference between the two samples. The results are provided in Table 9.

Second, we compared the distributions of participants' demographic and baseline health-related characteristics among all the dropout samples across the five experimental groups. We then conducted one-way ANOVA test and could not reject the null hypothesis that all the five groups are from the same sample distribution. The detailed results are provided in Table 10.

The results from the above two tests show that although dropout rate is non-negligible (~15%) in our study, the distribution of dropout samples remains quite consistent with that of the eligible samples, and moreover, the distribution of dropout samples remains quite consistent across the five experimental groups (i.e., missing data at random). Therefore, while we acknowledge this fact as one potential data limitation in our study, we are more confident that it is not a serious concern in affecting our results.

**7.3 Validity Check for Self-Reported Data**

Because medical information is sensitive, the accuracy of self-reported data is important for the validity of the results. We have validated our data using a multi-pronged approach.

First, our mHealth app partner provided an internal (full-time) medical expert team who helped review and validated the information about our experimental participants during the entire experimental period. In particular, as part of the risk assessment function, the internal medical team will communicate with each patient in person (mostly through phone calls) at least once every three months to carefully go over the historical (self-reported) records and the corresponding algorithm-generated diabetes risk score with the patient to better explain and validate the results. This validity check was done for all patients in C2, T1, T2 and T3 groups (who had access to the entire app functions through either PC or mobile devices). For patients in C1 group (who did not have access to the app), we checked the validity of their self-reported information during the surveys. In particular, during the phone calls we went through all the self-reported glucose values with them and validate their answers in person.

Second, during pre- and post-treatment surveys we purposely asked the same set of questions regarding the demographics and historical medical conditions. This to some extent helped cross validate the accuracy of the information (i.e., it is less likely a person will remember exactly what he/she said 8 months ago if that was a lie). Besides, the surveys were conducted through phone calls. The spontaneous in-person conversation also helped our researchers to spot anything suspicious (e.g., an obvious lie).

Third, from an experimental design perspective, because our participants were fully randomized into the experimental groups, even if any potential noise might exist in the individual data, such noise effect would be minor and likely to cancel out across the experimental groups due to randomization.

Therefore, based on the above efforts, we are confident about the validity of our data and the accuracy of our final results.

## 8. Conclusion and Future Directions

In this paper, we have examined the mHealth ecosystem and its health and economic impacts on diabetes patient outcomes. To achieve our goal, we partnered with a firm in Asia that provides the nation's largest mobile health app platform that specializes in diabetes care. We have designed and implemented a large-scale randomized field experiment based on unique observations from diabetes patients over several months.

Our research demonstrates the adoption of the mHealth platform has a statistically significant impact on improving patients' dietary and life style, leading to a reduction in patients' blood glucose, hospital visits, and medical expenses over time. We also find heterogeneous effects between personalized and non-personalized messages. Interestingly, non-personalized mobile messages with general diabetes-care guidance demonstrate a stronger impact on patient engagement with the app, behavior and life style change, and health improvement. While personalized mobile messages show a smaller impact on the above, they are more likely to encourage a substitution of offline or in-person interactions with telehealth. Our study indicates the mHealth apps have great potential in improving patients' health outcomes, by assisting them with behavior modification and chronic disease self-management. It also provides important insights into the design of such mHealth platform in transforming healthcare by substituting offline physician visits with telehealth and telemedicine.

On a broader note, our research significantly improves our understanding of human behavior and interactions with smart and connected mHealth platforms, and broadly in the consumer Internet of Things (IOT). Digital health platform infrastructures are often the manifestation of complex technological and social systems (Eisenmann et al. 2011) and can have profound implications on social and economic transactions. However, how humans interact with the mHealth infrastructures is not as well understood. Our study provides important managerial insights on issues that may influence individuals' sustained engagement with mobile and wearable technologies, health and wellness, adherence to treatment and wellness regimens, and patient welfare. It improves our understanding of the key mechanisms that drive individual health and wellness behavior and lifestyle changes through mobile technologies. Finally, it provides critical policy implications regarding the design of smart digital health platforms through effective, sustained usage of these emerging technologies.

Our paper has some limitations, which can serve as fruitful areas for future research. First, in our data sample, the majority of the diabetes patients have type 2 diabetes (approximately 98%). Although type 2 diabetes accounts for approximately 90% to 95% of all diagnosed cases of diabetes,[17] an examination of the mHealth impact on other types of diabetes with a larger sample in future would be useful.

---

[17] http://www.healthline.com/health/type-2-diabetes/statistics

Second, in this current study, we have evaluated the mHealth app as a bundle of all the major functions. However, breaking down the overall application into different functional components (e.g., Behavior Tracking, Risk Assessment and Personalized Solution, Q&A, and Patient Community) and examining the health and economic impacts from each of them separately would be interesting. It would also be interesting to examine the impact of personalized reminders when the messages are positive vs negative as people may react differently to positive vs negative sentiments.

Third, in this paper, we have not considered the potential impact related to the textual content of patient-physician communications, mainly because of potential privacy concerns blocking access to the textual content of the personal communications. However, based on our conversation with the platform, we believe these patient-physician communications are highly professional and provide similar quality in medical guidance. In addition, in our analyses, we are able to control the frequency of the patient-physician communications.

Finally, our research focuses on the context of diabetes-care management. The methodologies and insights have the potential to be generalized to other chronic-disease and wellness-care contexts. However, examining other medical scenarios to compare the relationship and heterogeneity in the impact of the mHealth platform on patient behavior and outcomes under different healthcare contexts would be interesting and important for future research.

## Table 1. Summary Statistics of Main Variables

| Variable | Description | Mean | Std. | Min | Max |
|---|---|---|---|---|---|
| C1 | Dummy for control group 1 | 0.15 | 0.33 | 0 | 1 |
| C2 | Dummy for control group 2 (Web) | 0.20 | 0.40 | 0 | 1 |
| T1 | Dummy for treatment group 1 | 0.21 | 0.42 | 0 | 1 |
| T2 | Dummy for treatment group 2 | 0.22 | 0.43 | 0 | 1 |
| T3 | Dummy for treatment group 3 | 0.22 | 0.43 | 0 | 1 |
| **Patient Demographics** | | | | | |
| **Male** | Whether the patient is male | 0.65 | 0.47 | 0 | 1 |
| **Age** | Numerical value of age | 55.17 | 8.91 | 23 | 72 |
| **Age_30** | Dummy for age group <30 | 0.24 | 0.43 | 0 | 1 |
| **Age_30_40** | Dummy for age group 31-40 | 0.30 | 0.46 | 0 | 1 |
| **Age_41_60** | Dummy for age group 41-60 | 0.39 | 0.49 | 0 | 1 |
| **Age_60** | Dummy for age group >60 | 0.06 | 0.24 | 0 | 1 |
| **Married** | Whether the patient is married | 0.83 | 0.39 | 0 | 1 |
| **Income** | Numerical value of income ($, annual) | 76827.27 | 12258.67 | 29630 | 234524 |
| Income_50K | Dummy for income < 50K | 0.24 | 0.43 | 0 | 1 |
| Income_50_100K | Dummy for income 50-100K | 0.66 | 0.49 | 0 | 1 |
| Income_100_200K | Dummy for income 100,001-200K | 0.09 | 0.29 | 0 | 1 |
| Income_200K | Dummy for income >200K | 0.01 | 0.11 | 0 | 1 |
| **Patient Prior Conditions** | | | | | |
| **Pre-meal Glucose** | Prior (most recent) pre-meal glucose value | 7.23 | 1.83 | 3.2 | 18 |
| **Post-meal Glucose** | Prior (most recent) post-meal glucose value | 9.86 | 4.36 | 4.2 | 30.7 |
| **Hemoglobin** | Most recent glycated hemoglobin | 6.72 | 1.98 | 4.6 | 35 |
| **Complication** | Whether there is a complication | 0.19 | 0.39 | 0 | 1 |
| **Smoking** | Whether the patient is a smoker | 0.09 | 0.28 | 0 | 1 |
| **Drinking** | Whether the patient drinks >140ml alcohol per week | 0.08 | 0.25 | 0 | 1 |
| **Pregnant** | Whether the patient is pregnant | 0.01 | 0.12 | 0 | 1 |
| **Other Major Disease** | Whether the patient has other major diseases | 0.01 | 0.11 | 0 | 1 |
| **Type 2 Diabetes** | Whether the patient has Type 2 diabetes | 0.98 | 0.12 | 0 | 1 |
| **Type 1 Diabetes** | Whether the patient has Type 1 diabetes | 0.01 | 0.11 | 0 | 1 |
| **Gestational Diabetes** | Whether the patient has gestational diabetes | 0.01 | 0.12 | 0 | 1 |
| **Diabetes Age** | Year(s) since diabetes was first diagnosed | 5.40 | 5.14 | 0 | 28 |
| **Patient Health Outcomes** | | | | | |
| **Uploaded Glucose** | Patient self-uploaded real-time glucose (overall) | 7.18 | 2.07 | 3.1 | 34.3 |
| (Pre-meal) | Patient self-uploaded real-time glucose (pre-meal) | 6.47 | 1.70 | 3.1 | 29.1 |
| (Post-meal) | Patient self-uploaded real-time glucose (post-meal) | 8.17 | 2.18 | 3.9 | 34.3 |
| **Upload_Morning** | Whether uploading time is morning | 0.36 | 0.48 | 0 | 1 |
| **Upload_Afternoon** | Whether uploading time is afternoon | 0.15 | 0.36 | 0 | 1 |
| **Upload_Night** | Whether uploading time is night | 0.49 | 0.50 | 0 | 1 |
| **Hospital Visits** | Number of hospital visits related to diabetes during the last 3 months | 2.64 | 6.69 | 0 | 12 |
| **Medical Spending** | Amount of medical spending related to diabetes during the last 3 months ($) | 57.14 | 63.49 | 20 | 1587.30 |
| **Patient Activities** | | | | | |
| **Daily #Steps** | Number of steps walked per day | 3597.82 | 5123.67 | 1021 | 49926 |
| **Daily Exercise Time** | Daily exercise time (minutes) | 55.26 | 62.15 | 0 | 269.01 |

| | | | | |
|---|---|---|---|---|
| **Daily Exercise Calorie** | Daily calories burned through exercise | 330.19 | 372.40 | 0 | 1720 |
| **Daily Food Calories** | Amount of calories consumed per day | 1090.59 | 169.11 | 438 | 2647 |
| **Daily Sleeping Length** | Total daily sleeping time (minutes) | 559.28 | 196.60 | 198 | 1380 |
| **Weekly Late Night Sleep** | # Nights per week when go to bed after 11pm | 1.97 | 4.21 | 0 | 7 |
| colspan="6" | Patient App Usage |
| **Daily #Opening App** | Daily frequency of opening the app | 1.14 | 0.43 | 0 | 6 |
| **Daily #Activity Logs** | Daily frequency of activities documented through app | 1.32 | 4.65 | 0 | 34 |
| **Weekly #Communications** | Weekly # of in-app communications with physicians | 1.94 | 3.26 | 0 | 9 |
| **Weekly Loyalty Rewards** | Weekly loyalty rewards earned | 19.43 | 241.32 | 0 | 35000 |
| **Weekly In-app Shopping** | Weekly in-app shopping for health products ($) | 25.97 | 180.60 | 0 | 2541.43 |

#Observations on Uploaded Glucose Values: n=9,251, #patients n=1,070.
#Observations on Patient Activities and App Usage: n=55,359, #patients n=1,070.
Data Period: May 2015 – July 2016.

### Table 2. Randomization Check – Demographic and Baseline Characteristics across 5 Groups

| Variable | C1 (n=156) | C2 (n=209) | T1 (n=230) | T2 (n=234) | T3 (n=241) | ANOVA |
|---|---|---|---|---|---|---|
| **Age** | | | | | | |
| <30 | 23% | 22% | 24% | 21% | 24% | p>0.05 |
| 30-40 | 31% | 29% | 26% | 23% | 21% | p>0.05 |
| 41-60 | 40% | 42% | 45% | 51% | 48% | p>0.05 |
| >60 | 6% | 6% | 5% | 5% | 6% | p>0.05 |
| **Gender** | | | | | | |
| Male | 65% | 64% | 65% | 67% | 66% | p>0.05 |
| Female | 35% | 36% | 34% | 34% | 35% | p>0.05 |
| **Married** | 82% | 77% | 90% | 90% | 86% | p>0.05 |
| **Income** ($, annual) | | | | | | |
| <50K | 26% | 25% | 27% | 24% | 27% | p>0.05 |
| 50-100K | 66% | 65% | 62% | 65% | 64% | p>0.05 |
| 100,001-200K | 7% | 8% | 10% | 10% | 8% | p>0.05 |
| >200K | 1% | 1% | 1% | 1% | 1% | p>0.05 |
| **Baseline Condition** | | | | | | |
| Pre-meal Glucose | 7.11 | 7.04 | 6.90 | 7.13 | 6.95 | p>0.05 |
| Post-meal Glucose | 8.43 | 8.59 | 8.44 | 8.38 | 8.68 | p>0.05 |
| Glycated Hemoglobin | 7.03 | 6.98 | 6.60 | 6.67 | 6.82 | p>0.05 |
| Complication | 19% | 20% | 17% | 18% | 16% | p>0.05 |
| Smoking | 8% | 9% | 10% | 9% | 9% | p>0.05 |
| Type 2 Diabetes | 98% | 96% | 97% | 96% | 96% | p>0.05 |
| Type 1 Diabetes | 1% | 2% | 2% | 2% | 2% | p>0.05 |
| Gestational Diabetes | 1% | 1% | 2% | 2% | 2% | p>0.05 |
| Diabetes Age | 5.41 | 5.32 | 5.46 | 5.42 | 5.36 | p>0.05 |

*Note:* Data are in percentage or mean value. Percentages do not add up to 100% in some cases because of rounding. The majority of our patient samples belong to type 2 diabetes, which is the main focus of our study. Income is adjusted based on the local cost of living.
To better control for the potential variation in the patient-level characteristics, we also included all these variables in our primary analyses as control variables.

### Table 3. Results from the Group Mean Analysis

| Treatment Group | Diff-Glucose | Diff-Hemoglobin | Diff-Hospital Visits (Recent 3Mons) | Diff-Spending (Recent 3Mons, USD) |
|---|---|---|---|---|
| C1 (n=156) | -0.0287 | -0.0143 | -0.0283 | -0.95 |
| C2 (n=209) | -0.5173 | -0.1967 | -0.0568 | -5.70 |
| T1 (n=230) | -0.6291 | -1.0316 | -0.1208 | -8.55 |
| T2 (n=234) | -0.6790 | -1.1612 | -0.1393 | -11.55 |
| T3 (n=241) | -0.5746 | -0.9405 | -0.2264 | -31.00 |

*Note*: Values are calculated based on the difference between the two surveys (post-treatment value minus pre-treatment value). Glucose value is calculated based on an average across all glucose types. (**Pairwise t-Test** was conducted to test the pairwise difference between each two experimental groups for each of the four health outcome variables. The null hypothesis was rejected at **P<0.05** for each comparison.)

### Table 4. Estimation Results on Glucose Change from the Primary Diff-in-Diff Models

| Variables | Coef. (Std. Err.)$^{I}$ | Coef. (Std. Err.)$^{II}$ | Coef. (Std. Err.)$^{III}$ | Coef. (Std. Err.)$^{IV}$ |
|---|---|---|---|---|
| **Treatment Effect ($\beta_3$)** | | | | |
| $C2 \times Time_t$ | -0.3448** (0.1804) | -0.4606***(0.1805) | -0.4105** (0.1819) | -0.5106** (0.2059) |
| $T1 \times Time_t$ | -0.4107***(0.1553) | -0.4871***(0.1588) | -0.4642***(0.1589) | -0.5733***(0.1832) |
| $T2 \times Time_t$ | -0.4589***(0.1551) | -0.5327***(0.1565) | -0.4588***(0.1587) | -0.6170***(0.1816) |
| $T3 \times Time_t$ | -0.3753** (0.1506) | -0.4669***(0.1520) | -0.4243** (0.1531) | -0.5408** (0.1802) |
| $C2$ ($\beta_1$) | 1.4013 (1.5766) | 3.2889 (2.6396) | 1.5622 (0.9973) | 4.7363 (3.4386) |
| $T1$ ($\beta_1$) | 0.8605 (0.6704) | 0.8829 (0.6837) | 0.8565 (0.6912) | 1.1794 (1.0350) |
| $T2$ ($\beta_1$) | 0.8282 (0.6747) | 0.9042 (0.6893) | 0.9756 (0.6919) | 1.1432* (0.6361) |
| $T3$ ($\beta_1$) | 0.9583 (0.6784) | 0.9424 (0.6893) | 1.0193 (0.6893) | 1.2649* (0.6347) |
| $Time_t$ ($\beta_2$) | 0.3095** (0.1528) | 0.3920***(0.1545) | 0.3674** (0.1559) | 0.4755***(0.1822) |
| $Intercept$ ($\beta_0$) | 13.3714***(0.9649) | 11.4988***(1.0279) | 11.8268***(0.8336) | 10.5798***(1.8447) |
| **Patient-Specific Control Variables ($X_i$)** | | | | |
| Age, Married, Gender, Income, Prior Glucose, Prior Hemoglobin, Prior Medication, Other Disease, Complication, Smoking/Drinking, Pregnant, Diabetes Type, Interaction with Physicians. | Yes | Yes | ---- | ---- |
| **Patient-Time-Specific Control Variables ($C_{it}$)** | | | | |
| Diabetes Age, Uploaded Glucose Type, Upload Time/Day/Month, Daily Exercise (#Steps), Daily App Usage (daily frequency of opening the app, daily frequency of documenting activity logs, weekly frequency of communications, weekly loyalty rewards and other in-app engagement like shopping). | Yes | ---- | Yes | ---- |

*Note:* * p<0.1, ** p<0.05, *** p<0.01. Errors are clustered at the experimental group level. Age and Income are in log form. Models I~ IV include different sets of control variables. #patients=1,070, #observations=9,251.

Table 5. Estimation Results on Glucose Change Using Diff-in-Diff Model with Patient-Level Fixed Effects

| Variables | Coef. (Std. Err.)[V] | Coef. (Std. Err.)[VI] |
|---|---|---|
| **Treatment Effect ($\beta_2$)** | | |
| $C2 \times Time_t$ | -0.3327** (0.1704) | -0.4267** (0.1977) |
| $T1 \times Time_t$ | -0.3461** (0.1795) | -0.4349** (0.1945) |
| $T2 \times Time_t$ | -0.4909*** (0.1752) | -0.5172*** (0.1703) |
| $T3 \times Time_t$ | -0.4430** (0.1951) | -0.4873** (0.1944) |
| $Time_t$ ($\beta_1$) | 0.3557** (0.1732) | 0.3936** (0.1572) |
| $Intercept$ ($\beta_0$) | 10.1937*** (1.7438) | 7.3579*** (1.1258) |
| **Patient-Time-Specific Control Variables ($C_{it}$)** | | |
| Diabetes Age, Uploaded Glucose Type, Upload Time/Day/Month, Daily Exercise (#Steps), Daily App Usage (daily frequency of opening the app, daily frequency of documenting activity logs, weekly frequency of communications, weekly loyalty rewards and other in-app engagement like shopping). | Yes | ---- |

Note: * p<0.1, ** p<0.05, *** p<0.01. Errors are clustered at experimental group level. Models I~ IV include different sets of control variables. #patients=1,070, #observations=9,251.

Table 6a. Estimation Results on Patient Activities Using Diff-in-Diff Model with Patient Fixed Effects

| Variables | Daily Food Calories Intake Coef. (Std. Err.)[A1] | Daily Exercise Time Coef. (Std. Err.)[A2] | Daily Exercise Calories Coef. (Std. Err.)[A3] | Daily #Steps Walked Coef. (Std. Err.)[A4] | Daily Sleeping Length Coef. (Std. Err.)[A5] | Weekly Freq of Late Night Sleep Coef. (Std. Err.)[A6] |
|---|---|---|---|---|---|---|
| **Treatment Effect** | | | | | | |
| $C2 \times Time_t$ | ---- | ---- | ---- | ---- | ---- | ---- |
| $T1 \times Time_t$ | -0.1803*** (0.0258) | 0.0642** (0.0258) | 0.0626** (0.0241) | 0.0326** (0.0161) | 0.1417** (0.0710) | 0.0332 (0.0273) |
| $T2 \times Time_t$ | -0.1854*** (0.0227) | 0.0688*** (0.0262) | 0.0681** (0.0265) | 0.0434*** (0.0168) | 0.1814** (0.0719) | 0.0306 (0.0283) |
| $T3 \times Time_t$ | -0.1843*** (0.0223) | 0.0457* (0.0250) | 0.0472** (0.0233) | 0.0218* (0.0165) | 0.1019* (0.0691) | 0.0423* (0.0278) |
| $Time_t$ ($\beta_1$) | 0.0197 (0.0234) | -0.0331 (0.0274) | -0.0135 (0.0221) | 0.0102 (0.0162) | -0.0298 (0.0575) | -0.0115 (0.0245) |
| $Intercept$ ($\beta_0$) | 2.1149*** (0.0206) | 1.1162*** (0.0229) | 1.4566*** (0.0262) | 1.3729*** (0.0175) | 1.7192*** (0.0575) | 0.4192*** (0.0302) |

Note: * p<0.1, ** p<0.05, *** p<0.01. Errors are clustered at experimental group level.
Models A1~A6 correspond to the following user activity outcome variables: A1 – (log) Daily food calories intake, A2 – (log) Daily exercise time (mins), A3 – (log) Daily exercise calories, A4 – (log) Daily #steps walked, A5 – (log) Daily sleeping length (mins), A6 – #Nights per week when the patient went to sleep later than 11pm.
#Patients=1,070, #Observations=55,359

**Table 6b. Estimation Results on Patient Activities Using Diff-in-Diff Model with Patient Fixed Effects and Additional Patient-Time-Specific Control Variables**

| Variables | Daily Food Calories Intake Coef. (Std. Err.)A1 | Daily Exercise Time Coef. (Std. Err.)A2 | Daily Exercise Calories Coef. (Std. Err.)A3 | Daily #Steps Walked Coef. (Std. Err.)A4 | Daily Sleeping Length Coef. (Std. Err.)A5 | Weekly Freq of Late Night Sleep Coef. (Std. Err.)A6 |
|---|---|---|---|---|---|---|
| **Treatment Effect** | | | | | | |
| $C2 \times Time_t$ | ---- | ---- | ---- | ---- | ---- | ---- |
| $T1 \times Time_t$ | -0.1677*** | 0.0597** | 0.0601** | 0.0313** | 0.1399** | 0.0320 |
| | (0.0243) | (0.0252) | (0.0240) | (0.0162) | (0.0711) | (0.0275) |
| $T2 \times Time_t$ | -0.1724*** | 0.0602*** | 0.0692** | 0.0452*** | 0.1826** | 0.0329 |
| | (0.0202) | (0.0268) | (0.0262) | (0.0169) | (0.0722) | (0.0288) |
| $T3 \times Time_t$ | -0.1714*** | 0.0431* | 0.0455** | 0.0266* | 0.1112* | 0.0443* |
| | (0.0204) | (0.0252) | (0.0235) | (0.0168) | (0.0694) | (0.0279) |
| $Time_t\ (\beta_1)$ | 0.0183 | -0.0348 | -0.0132 | 0.0101 | -0.0275 | -0.0122 |
| | (0.0219) | (0.0271) | (0.0222) | (0.0163) | (0.0577) | (0.0247) |
| $Intercept\ (\beta_0)$ | 1.9668*** | 1.1198*** | 1.4574*** | 1.3818*** | 1.7158*** | 0.4432*** |
| | (0.0198) | (0.0226) | (0.0267) | (0.0177) | (0.0578) | (0.0305) |
| **Patient-Time-Specific Control Variables:** | | | | | | |
| Diabetes Age, Daily App Usage (daily frequency of opening the app, daily frequency of documenting activity logs, weekly frequency of communications, weekly loyalty rewards and other in-app engagement like shopping). | | | | | | |
| | Yes | Yes | Yes | Yes | Yes | Yes |

*Note:* * $p<0.1$, ** $p<0.05$, *** $p<0.01$. Errors are clustered at experimental group level.
Models A1~A6 correspond to the following user activity outcome variables: A1 – (log) Daily food calories intake, A2 – (log) Daily exercise time (mins), A3 – (log) Daily exercise calories, A4 – (log) Daily #steps walked, A5 – (log) Daily sleeping length (mins), A6 – #Nights per week when the patient went to sleep later than 11pm. #Patients=1,070, #Observations=55,359

Table 7. Estimation Results on App Usage Using Diff-in-Diff Model with Patient Fixed Effects

| Variables | Daily Freq of Opening App Coef. (Std. Err.)U1 | Daily Freq of Documenting Activity Logs Coef. (Std. Err.)U2 | Weekly Freq of Communi-cations Coef. (Std. Err.)U3 | Weekly Loyalty Rewards Coef. (Std. Err.)U4 | Weekly in-app Shopping (Total Purchase $) Coef. (Std. Err.)U5 |
|---|---|---|---|---|---|
| **Treatment Effect** | | | | | |
| $C2 \times Time_t$ | ---- | ---- | ---- | ---- | ---- |
| $T1 \times Time_t$ | 0.1808*** | 0.2734*** | 0.0812*** | 0.2678* | -0.0023 |
|  | (0.0190) | (0.0873) | (0.0059) | (0.1289) | (0.2652) |
| $T2 \times Time_t$ | 0.1951*** | 0.2965*** | 0.0872*** | 0.2863* | 0.0375 |
|  | (0.0190) | (0.0881) | (0.0058) | (0.1266) | (0.2879) |
| $T3 \times Time_t$ | 0.1243*** | 0.2160** | 0.0241*** | 0.1557** | -0.1127 |
|  | (0.0185) | (0.0893) | (0.0049) | (0.1240) | (0.3201) |
| $Time_t (\beta_1)$ | -0.0401** | -0.1626** | -0.1009*** | -0.1906* | 0.0870 |
|  | (0.0176) | (0.0797) | (0.0051) | (0.1107) | (0.2035) |
| $Intercept (\beta_0)$ | 1.0187*** | 2.2540*** | 1.2167*** | 2.0932*** | 0.8096** |
|  | (0.0167) | (0.0802) | (0.0058) | (0.1224) | (0.3317) |

***Note:*** * $p<0.1$, ** $p<0.05$, *** $p<0.01$. Errors are clustered at experimental group level. Models U1~U5 correspond to the following app usage outcome variables: U1 – (log) Daily frequency of opening the mHealth app, U2 – (log) Daily frequency of documenting activities through the app, U3 – Weekly frequency of communications with medical experts, U4 – (log) Weekly loyalty rewards earned, U5 – (log) Weekly shopping total purchase ($).  #Patients=1,070, #Observations=55,359

Table 8. Subsample Analysis (Patients Recruited in May 2015)

| Treatment | Diff-Glucose | Diff-Hemoglobin | Diff-Hospital Visits (Recent 3Mons) | Diff-Spending (Recent 3Mons, USD) |
|---|---|---|---|---|
| C1 (n=49) | -0.0338 | -0.0149 | -0.0297 | -0.88 |
| C2 (n=63) | -0.5202 | -0.1943 | -0.0601 | -5.62 |
| T1 (n=57) | -0.6312 | -1.0307 | -0.1319 | -9.69 |
| T2 (n=64) | -0.6978 | -1.1588 | -0.1443 | -14.70 |
| T3 (n=52) | -0.5889 | -0.9576 | -0.2098 | -29.63 |

***Note:*** Values are calculated based on the difference between the two surveys (post-treatment value minus pre-treatment value). Glucose value is calculated based on an average across all glucose types. **P<0.05 (ANOVA)**

Table 9. Comparison of Main Variables between Eligible Samples and Dropout Samples

| | Eligible Samples | | Dropout Samples | | t-test |
|---|---|---|---|---|---|
| **Variable** | **Mean** | **Std.** | **Mean** | **Std.** | |
| **Male** | 0.65 | 0.47 | 0.63 | 0.49 | t = 1.39 (p>0.05) |
| **Age** | 55.17 | 8.91 | 54.01 | 8.82 | t = 1.93 (p>0.05) |
| **Married** | 0.83 | 0.39 | 0.79 | 0.35 | t = 1.65 (p>0.05) |
| **Income** | 76827.27 | 12258.67 | 75403.57 | 13179.28 | t = 1.62 (p>0.05) |
| **Pre-meal Glucose** | 7.23 | 1.83 | 7.32 | 1.92 | t = 1.61 (p>0.05) |
| **Post-meal Glucose** | 9.86 | 4.36 | 9.95 | 4.22 | t = 0.68 (p>0.05) |
| **Hemoglobin** | 6.72 | 1.98 | 6.81 | 1.87 | t = 1.49 (p>0.05) |
| **Diabetes Age** | 5.40 | 5.14 | 5.11 | 5.02 | t = 1.85 (p>0.05) |
| Eligible samples: #patients n=1,070. | | | Dropout Samples: #patients n=273 | | |

Table 10. Comparison of Main Variables among Dropout Samples across Five Experimental Groups

| Variable | C1 Mean | C2 Mean | T1 Mean | T2 Mean | T3 Mean | ANOVA |
|---|---|---|---|---|---|---|
| Male | 0.65 | 0.64 | 0.63 | 0.65 | 0.66 | p>0.05 |
| Age | 55.21 | 54.68 | 54.14 | 54.79 | 55.02 | p>0.05 |
| Married | 0.83 | 0.81 | 0.80 | 0.81 | 0.79 | p>0.05 |
| Income | 76809 | 76092 | 75631 | 75395 | 78909 | p>0.05 |
| Pre-meal Glucose | 7.21 | 7.27 | 7.32 | 7.24 | 7.36 | p>0.05 |
| Post-meal Glucose | 9.89 | 9.82 | 9.91 | 9.94 | 9.86 | p>0.05 |
| Hemoglobin | 6.78 | 6.72 | 6.80 | 6.83 | 6.79 | p>0.05 |
| Diabetes Age | 5.38 | 5.25 | 5.16 | 5.21 | 5.19 | p>0.05 |
| Sample Size: C1(n=97), C2(n=92), T1(n=23), T2(n=35), T3(n=26) | | | | | | |

Table 11. Estimation Results Using Simultaneous Equation Model – Exercise and Glucose

| Glucose Change (Post – Pre) | Coef. (Std. Err.)$^V$ |
|---|---|
| *Exercise Calories (log)* | -0.6935***(0.0974) |
| *Intercept* | 10.8099***(1.7367) |
| *Patient -Specific Control Variables* | |
| Age, Married, Gender, Income, Prior Glucose, Prior Hemoglobin, Prior Medication, Other Disease, Complication, Smoking/Drinking, Pregnant, Diabetes Type, Interaction with Physicians. Diabetes Age, average Daily App Usage (daily frequency of opening the app, daily frequency of documenting activity logs, weekly frequency of communications, weekly loyalty rewards and other in-app engagement like shopping). | Yes |
| **Exercise Calories (log)** | |
| *T1* | 0.2148***(0.0551) |
| *T2* | 0.2498** (0.1232) |
| *T3* | 0.1661***(0.0678) |
| *Intercept* | 5.0485***(0.1348) |
| *Patient -Specific Control Variables* | |
| Age, Married, Gender, Income, Prior Glucose, Prior Hemoglobin, Prior Medication, Other Disease, Complication, Smoking/Drinking, Pregnant, Diabetes Type, Interaction with Physicians. Diabetes Age, average Daily App Usage (daily frequency of opening the app, daily frequency of documenting activity logs, weekly frequency of communications, weekly loyalty rewards and other in-app engagement like shopping). | Yes |
| *Note:* * $p<0.1$, ** $p<0.05$, *** $p<0.01$. #patients=1,070, #observations=9,251. | |

**Table 12. Estimation Results Using Simultaneous Equation Model – Food Intake and Glucose**

| Glucose Change (Post – Pre) | Coef. (Std. Err.) |
|---|---|
| *Food Calories Intake (log)* | 0.3760***(0.1258) |
| *Intercept* | 7.5798***(1.4316) |
| *Patient -Specific Control Variables* | |
| Age, Married, Gender, Income, Prior Glucose, Prior Hemoglobin, Prior Medication, Other Disease, Complication, Smoking/Drinking, Pregnant, Diabetes Type, Interaction with Physicians. Diabetes Age, average Daily App Usage (daily frequency of opening the app, daily frequency of documenting activity logs, weekly frequency of communications, weekly loyalty rewards and other in-app engagement like shopping). | Yes |
| **Food Calories Intake (log)** | |
| *T1* | -0.5664***(0.1540) |
| *T2* | -0.8825***(0.1402) |
| *T3* | -0.2134***(0.0251) |
| *Intercept* | -6.9580***(2.0145) |
| *Patient -Specific Control Variables* | |
| Age, Married, Gender, Income, Prior Glucose, Prior Hemoglobin, Prior Medication, Other Disease, Complication, Smoking/Drinking, Pregnant, Diabetes Type, Interaction with Physicians. Diabetes Age, average Daily App Usage (daily frequency of opening the app, daily frequency of documenting activity logs, weekly frequency of communications, weekly loyalty rewards and other in-app engagement like shopping). | Yes |

*Note:* * $p<0.1$, ** $p<0.05$, *** $p<0.01$. #patients=1,070, #observations=9,251.

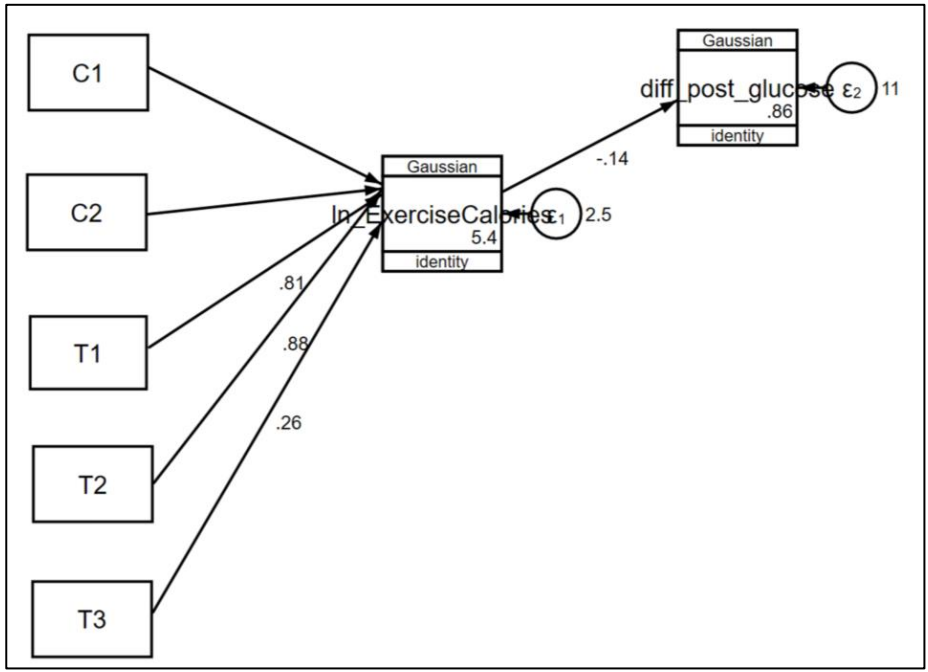

**Figure 1. Directed Acyclic Graph (DAG) to Test the Mediation Effect of Exercise on Post-Experiment Blood Glucose Change.**

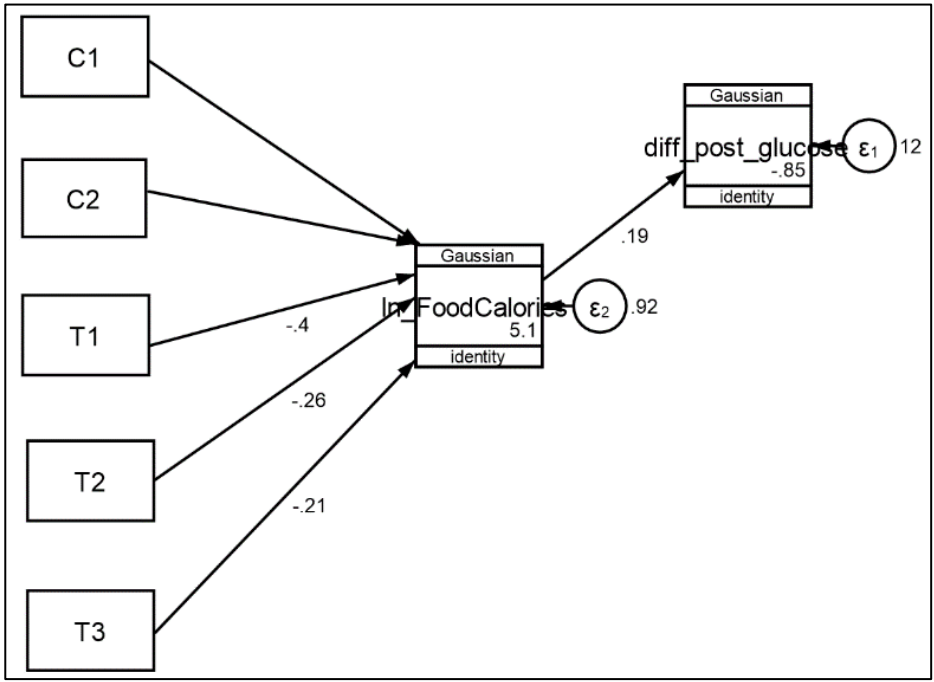

**Figure 2. Directed Acyclic Graph (DAG) to Test the Mediation Effect of Food Intake on Post-Experiment Blood Glucose Change**

**(Online Appendix)**
**Empowering Patients Using Smart Mobile Health Platforms: Evidence From A Randomized Field Experiment**

**Online Appendix A. Screenshots of Mobile/Web Interfaces**

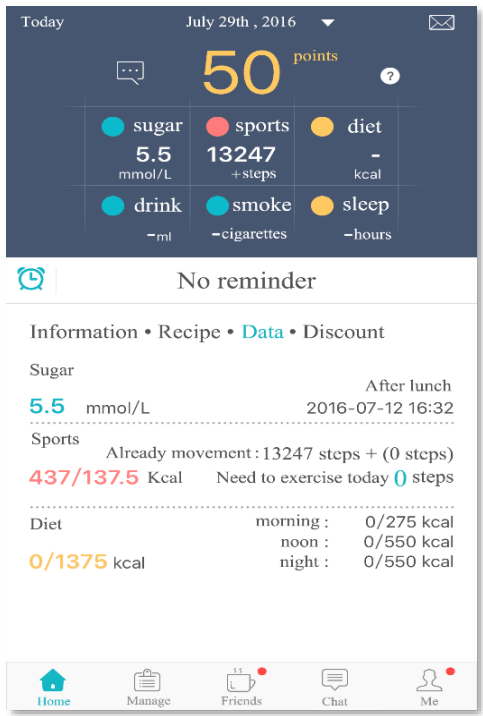

(1a) Overview of User Homepage

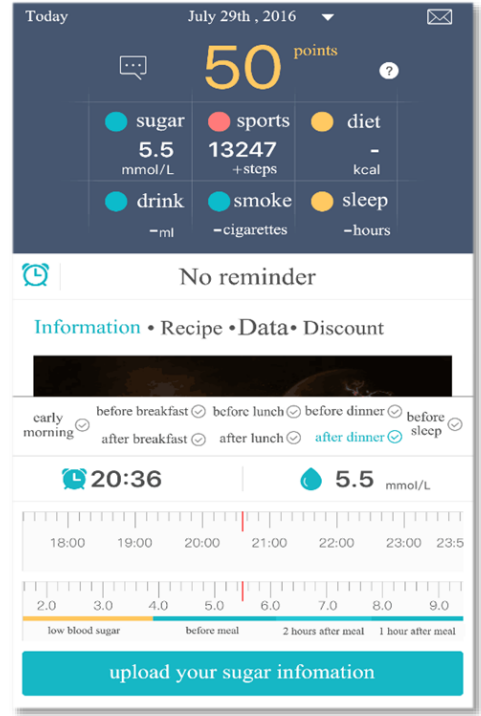

(1b) Adding a New Blood Glucose Value

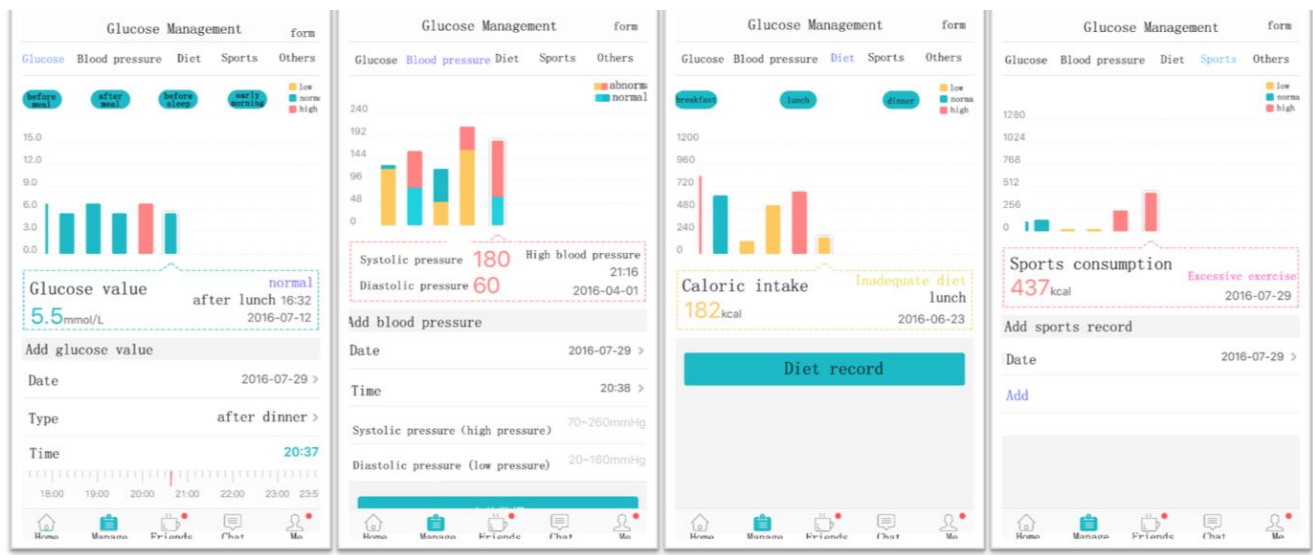

(1c) User Behavior Tracking over Time.
(from left to right: Glucose, Blood pressure, Diet, and Exercise (Sports))

**Figure A1. Screenshots of the Main App Functions**

**Figure A2. Screenshot of the Web Portal for Control Group C1 to Upload the Blood Glucose and Hemoglobin Values at the Beginning and End of the 3-month Treatment Period**

**Figure A3. Screenshots of the Behavior Recording Pages (Exercise and Diet)**

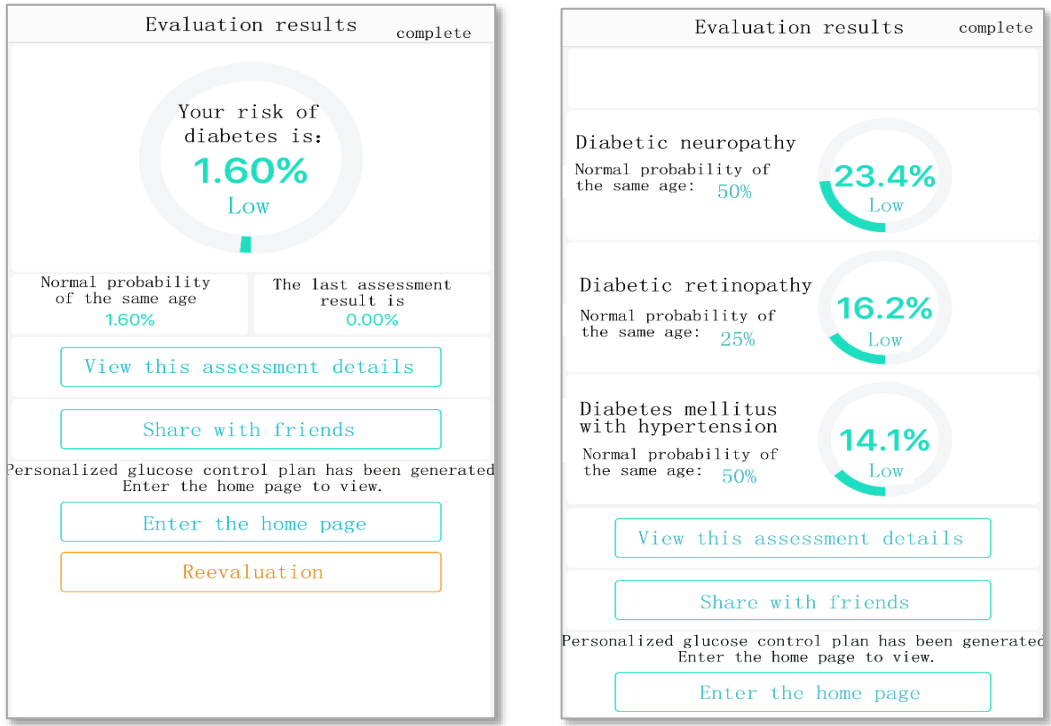

Figure A4. Screenshots of the Diabetes Risk Assessment Pages

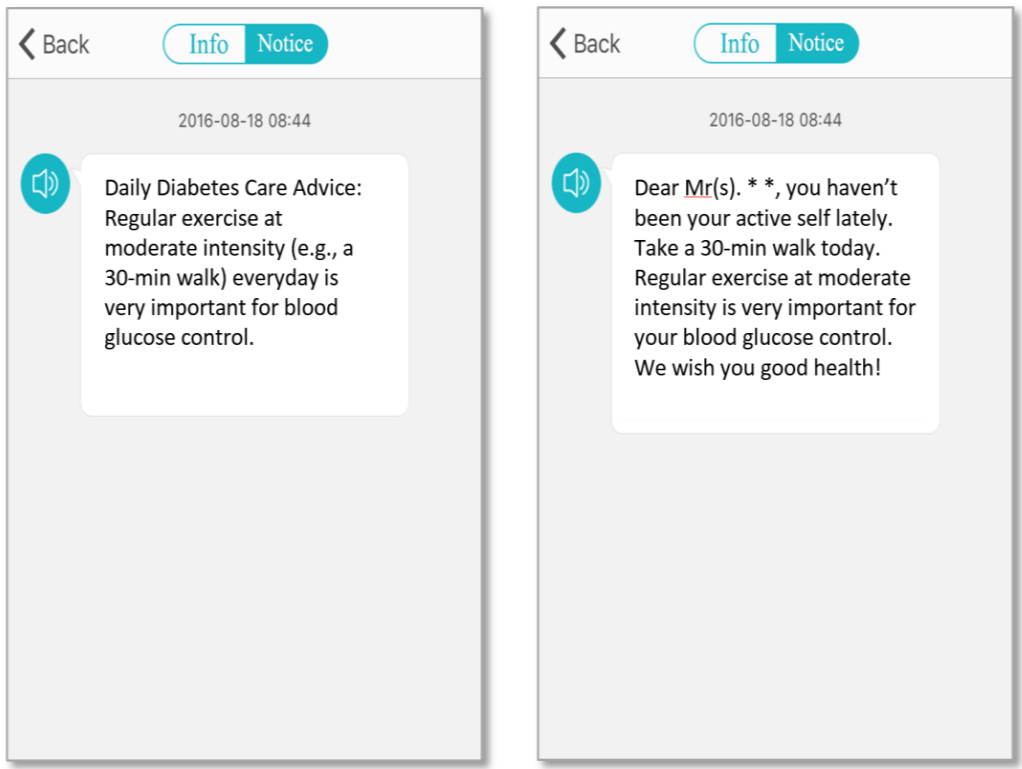

Figure A5. Screenshots of the Mobile Messages (Left: Non-personalized; Right: Personalized)

# Online Appendix B. Survey Questionnaires[18]

- **Pre-experiment questionnaire**

    1. What is your last two pre-meal blood glucose values (mmol/L)?
        A. 6.1-9.1
        B. 9.1-12.1
        C. 12.1-15.1
        D. Over 15.1
    2. What is your last two post-meal blood glucose values (mmol/L)?
        E. 6.1-9.1
        F. 9.1-12.1
        G. 12.1-15.1
        H. Over 15.1
    3. What's your age?
        A. Under 30 years old
        B. 30-40 years old
        C. 40-60 years old
        D. Over 60 years old
    4. How much do you spend monthly for your diabetes treatment?
        A. Less than 2000 RMB
        B. 2000-5000 RMB
        C. 5000-10000 RMB
        D. More than 10000 RMB
    5. How many types of medicine are you currently taking to treat diabetes?
        A. None
        B. 1-2
        C. 3-4
        D. 5 or more than 5
    6. How often do you test your blood sugar?
        A. Once a day
        B. 2-3 times per day
        C. Once every 2-3 days
        D. Once a week
        E. Others
    7. How do you evaluate your current diet?
        A. My diet is healthy and in line with the dietary requirements of people with diabetes.
        B. Quite regular, can eat three meals on time, can achieve less salt, less sugar, less oil
        C. Three meals a day, can be eaten on time, can try to achieve less salt, less sugar, less oil, but occasionally can't.
        D. Three meals a day, but cannot control foods that eat less salt, less sugar, less oil.
        E. Three meals are irregular, but can control less salt, less sugar, less oil
        F. Three meals are irregular, unable to control diet

---

[18] The questionnaires were translated from the original language into English for readability.

8. How do you think about healthy diet? (You can choose multiple options)
    A. Diet differentiation, eat more grains
    B. More pure natural food
    C. More fruit and vegetable
    D. Health care products, such as vitamin tablets
    E. I have no idea about healthy diet
    F. Others
9. Which of the following descriptions are appropriate for your daily workout? ( You can choose multiple options)
    A. I exercise lightly every day, like walking
    B. I participate in fitness activities every day, such as running, playing Tai Chi, square dance, etc.
    C. I rarely participate in physical exercise, I rarely go out.
    D. I usually do high-intensity exercises, such as weight-bearing anaerobic exercise, and occasionally mild exercise, such as walking and playing Tai Chi.
    E. I usually do mild exercise and occasionally do high-intensity exercises.
    F. I don't do any exercise.
10. Do you feel that your current exercise situation is conducive to the recovery of diabetes?
    A. Obvious effect
    B. General effect
    C. Almost no effect
    D. No idea
11. If jogging is good for your body every day, you can stick to it under the following conditions:
    A. If someone reminds you to jog every day
    B. If someone encourage you to jog every day
    C. If someone is running around every day
    D. Anyway, it's hard to stick to
12.  Which kind of emotions do you often have?[19]
    A. Joyful and happy
    B. Feeling depressed
    C. Anxiety and depression
    D. Peaceful
13. Do you purchase and consume sugar-free health food for diabetics?
    A. Long-term consumption, regular purchase, fixed purchase location
    B. Occasionally eat, occasionally purchased, no fixed place to buy
    C. Seldom eat, there are patients recommended to try, there is no fixed place to buy
    D. Do not trust such products, think that you can stick to the kiln and diet, do not buy
14. Do you have the confidence to beat diabetes?
    A. Very confident
    B. General Confident
    C. Confident, but think it is hard

---

[19] we asked the question about emotions mainly to screen participants who had unstable emotional status. We didn't find any such case, so we didn't include this variable in our main analysis.

D. Lack of confidence, but willing to try
   E. Lack of confidence, barely maintain the status
15. Have you ever participated in a diabetes rehabilitation program or a similar health management program?
    A. Yes, I have
    B. No, I have not
    C. No, but heard about that
16. What is your highest concern in health management? (You can choose multiple options )
    A. Regular medical examination service
    B. Personal health record establishment and management
    C. Self-monitoring
    D. Private doctor service
    E. Personalized health management
    F. Health guidance, lifestyle intervention and adjustment
    G. Lecture, salon, party about heath management
    H. Personal consultation service
17. Do you Smoke?
    A. Yes
    B. No
18. What type of diabetes do you have?
    A. Type 1 diabetes
    B. Type 2 diabetes
    C. Gestational diabetes
19. How long do you have diabetes?
    ___________Years
20. Have you had any complication?
    A. Yes, please specify______________________
    B. No

- **Post-experiment questionnaire (end of the treatment period, and 5 months later)**

    1. What is your last two pre-meal blood glucose values (mmol/L)?
        I. 6.1-9.1
        J. 9.1-12.1
        K. 12.1-15.1
        L. Over 15.1
    2. What is your last two post-meal blood glucose values (mmol/L)?
        M. 6.1-9.1
        N. 9.1-12.1
        O. 12.1-15.1
        P. Over 15.1
    3. How many times did you visit the hospital during the last 3 months?
        A. None
        B. 1-3
        C. 3-6

D. 6-10
   E. More than 10
4. How many of these hospital visits were related to diabetes?
   A. 0
   B. 1-3
   C. 3-6
   D. 6-10
   E. More than 10
5. How much do you spend monthly for your diabetes treatment?
   E. Less than 2000 RMB
   F. 2000-5000 RMB
   G. 5000-10000 RMB
   H. More than 10000 RMB
6. What is your current solution to manage your blood sugar?
   A. Taking hypoglycemic drugs
   B. Insulin
   C. Diet management
   D. Sports management
   E. Daily life management (sufficient sleep, smoking cessation, alcohol restriction, etc.)
7. What kinds of drugs do you currently take? [Multiple choice questions]
   A. Mtformin
   B. Acarbose
   C. Insulin
   D. Glipizide
   E. Gliclazide
   F. Giclazone
   G. Rpaglinide
   H. Sitagliptin
   I. Others
8. How many times do you use the app every day? (only for treatment group)
   A. 0
   B. 1
   C. 2
   D. 3
   E. More than 3
9. Can you rate the app? (only for treatment group)
   A. 1 star
   B. 2 stars
   C. 3 stars
   D. 4 stars
   E. 5 stars

## Online Appendix C. Overview of Randomization and Sampling

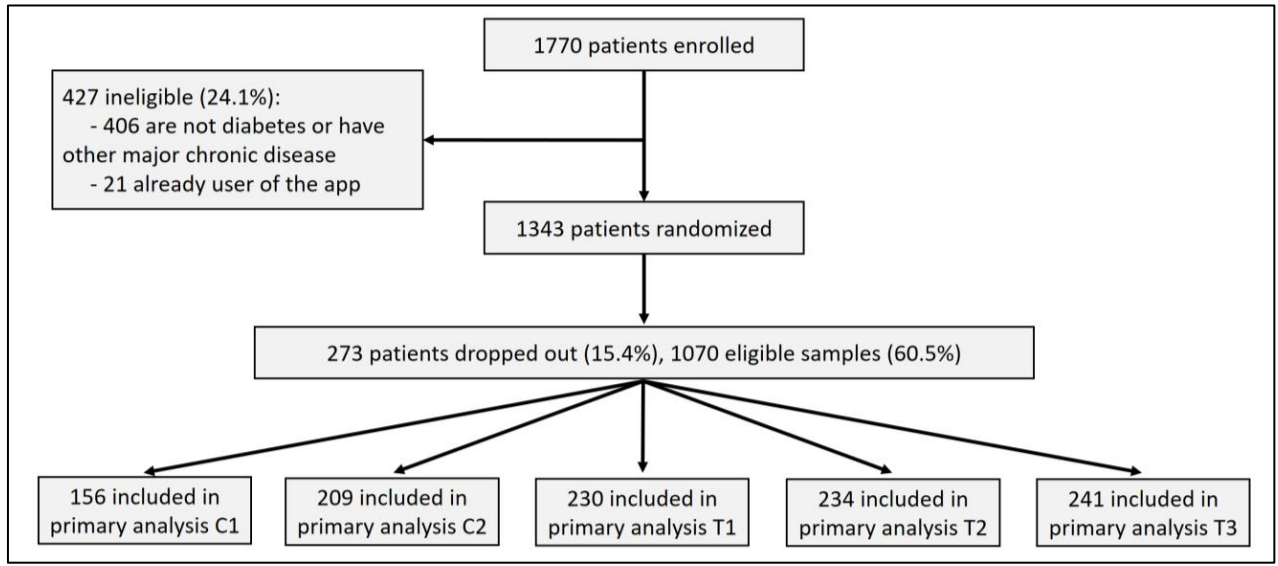

**Figure C1. Randomization and Sampling Procedure**

## Online Appendix D. Time Trends

We examined the overall time trends in each experimental group regarding the blood glucose change over time at the individual patient level. We plot the glucose value over time for each group in Figure D1.

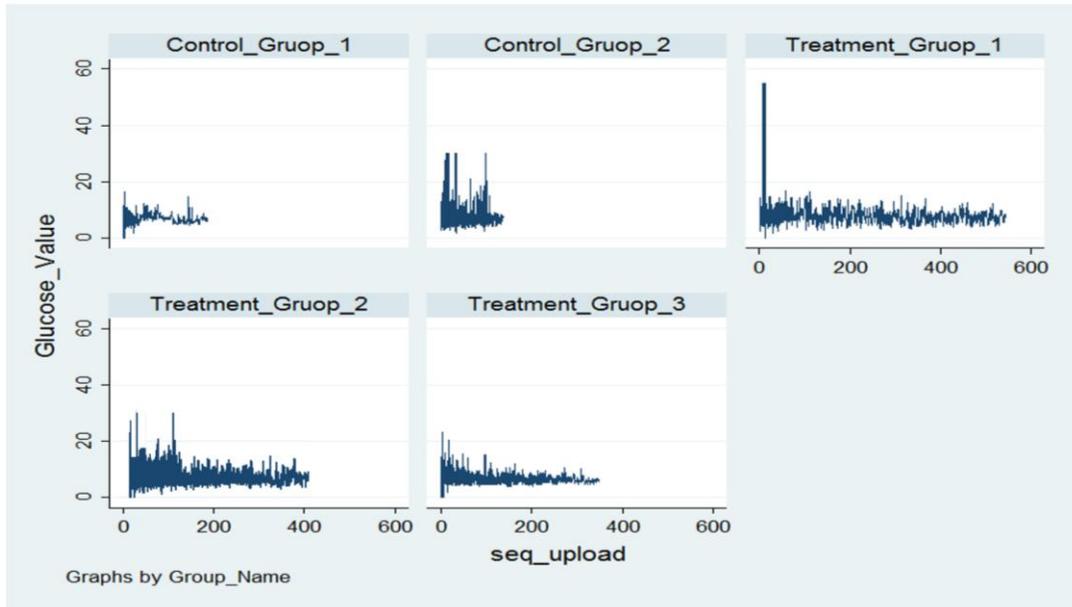

**Figure D1. Comparison of Time Trends for Blood Glucose Values over Time**

The Y-axis is the glucose value for each individual patient. The X-axis is the sequence number as the time indicator. We show the plots for both control groups and treatment groups at the individual level. From the time trend plots, we notice the three treatment groups on average uploaded more glucose values than the two control groups. This finding indicates a potential positive impact of mHealth in improving patient engagement with diabetes management. Furthermore, we see a noticeable downward trend over time in the three treatment groups compared to the two control groups. This finding suggests the mHealth platform seems to be able to help reduce patient glucose levels over time at the individual level. We also noticed an outlier in the T1 group at the very beginning, with a glucose value equal to 55. After consulting with the company and medical experts, we removed that sample from our primary model analysis.

## Online Appendix E. An Additional Follow-up Survey and Interview

To better understand the causal mechanisms of our findings, we have conducted a new round of survey and interview in September 2019.

**1. Summary of Survey Results.**

First, we sent a survey to app users from the three treatment groups (T1-T3) in our experiment. We approached these participants via WeChat groups the mobile app company maintained over years. There were 610 treatment participants (out of 705 total participants from T1-T3) in the WeChat groups. We received a total of 124 responses to our survey. The response rate was 124/610 = 20.3%.

The detailed survey questionnaire was provided in the end of this Appendix. We asked the participants three major questions: (1) What is your favorite function of the blood glucose management mobile app? (2) How did these functions help improve your health? (3) For participants in T3, what's your feedback towards the personalized text messages about medical guidance based on your personal exercise, diet and health status?

Based on the survey user responses, the most useful app function liked by the users is "*Learning about health knowledge* (68%)," followed by "*Health real-time tracking: blood glucose, exercise, drug, diet* (67%)," "*Personalized diabetes risk assessment* (61%)," "*Doctor consultation* (53%)," and "*Social network support* (48%)."

When being asked how these functions helped improve their health, a large majority of the users mentioned that the app provided them a way to better monitor and "quantify" their life and health in real time, hence they were able to better manage food intake and exercise. For example,

- *"The blood glucose, exercise and diet tracking function provides me with a tool for daily health monitoring and comparison."*
- *"The combination of my blood glucose level and exercise diet allows me to understand the relationship between them clearly, which motivates me to exercise more and eat healthier food."*
- *"Self-tracking of health status provides a quantitative basis in real time, and thus improves my health level."*
- *"Personalized recommendations for diet and exercise allow me to know exactly how many calories I should consume and how many I should burn."*

Regarding the personalized text message about medical guidance, users raised three major concerns: (1) Interruption and annoyingness (58%), (2) Avoidance towards negative information (54%), and (3) Privacy (53%). For example,

- *"Receiving personalized text messages frequently makes me feel interrupted and the user experience is terrible."*
- *"Sending personalized information frequently makes me feel terrible about my health, and too many negative emotions make me reluctant to try to change my health level."*
- *"It makes me feel my privacy is being violated."*

**2. Summary of Interview Results.**

Second, in addition to the survey responses, we have also conducted in-depth phone interviews with a randomly selected group of 7 experimental users from T3 treatment group. The main purpose of the interview was to further verify the survey responses, and meanwhile with a focus on why the personalized text messages did not work well. The detailed question design of the interview was provided in the end of this Appendix.

We found the responses from the interview were highly consistent with those from the survey. When the participants were asked what functions they liked the most, all of the 7 interview participants indicated that the real-time health tracking function provided them a way of better monitoring and managing their health. They (and their family members) have also gained professional health knowledge through using the app. For example,

- *"After using the app for some time, I have gained some health knowledge."*
- *"The real-time tracking and the long-term blood glucose change trend helped me judge whether the medication, exercise or diet was reasonable and provided me a reference."*
- *"You see, some people don't have a lot of knowledge about diabetes, right? People can gain some knowledge, and also have some reminders about their blood glucose management."*
- *"I found the health tracking function is most useful for me. In daily life, it can help monitor my blood sugar."*
- *"It can record my health data. I like this function most."*
- *"It certainly doesn't make sense to people who are not sick, but it definitely makes sense to us who are sick, because we need such an assistant to let us know how high our blood sugar is at any time. We are very concerned about this."*
- *"The most important one is the health knowledge. Because the patient's understanding of professional knowledge is still relatively inadequate. I have been learning a lot. Because I have had such instructions from the app, and my wife also has learned related knowledges and helped me."*

When the participants were asked whether they liked the personalized medical guidance via text messages and why, a majority of them (6 out of 7) indicated that they found these personalized text messages "too frequent," "annoying" and violating "privacy." For example,

- *"The personalized message was sent too often. It's better to have a longer interval."*
- *"I think it's a bit troublesome. I think the frequency of the personalize message was a little high."*
- *"I can accept the personalized advice, but you better not send the reminders so often."*
- *"If your blood glucose is in an expected range, you don't need to be disturbed. Just when it's abnormal (you can receive a personalized message intervention)"*
- *"It doesn't need to be reminded too often, just once every half a month. Because my blood sugar is now in a stable state."*
- *"It feels that it knows what I do and feels like I am being watched by others."*

Very interestingly, during our interview one of the participants explicitly mentioned his/her preference of a less personalized text message to avoid "being judged all the time by someone":

- *"Is it possible for you to **make these personalized guidance text messages sound less personal, but instead more systematic – like the ones automatically sent by the system, not humans?** That would make me feel less stressful. Not like being judged by someone all the time, but simply like having an alarm clock."*

This is highly consistent with our previous finding that frequent personalized messages might cause the patients to feel increased control and judgment. They might cause patients to feel pressured or coerced by intrapsychic or interpersonal forces. Such lack of choiceful and volitional feeling can lead to loss of autonomy and self-motivation (Deci and Ryan 1985, 2000). It in turn can lead to a significant decrease in patient intrinsic motivation of disease self-management and a lower health outcome.

In summary, our additional analyses from the new survey and interview demonstrated high consistency to our previous results. They provided richer causal evidence to our findings. We found that the positive impact of mobile health app is largely due to the real-time tracking and health monitoring functions provided by the app. Such functions can help patients with better self-educating, self-monitoring, and self-managing their own health. Besides, users raised three major concerns towards personalized health guidance via text messages: (1) Interruption and annoyingness, (2) Avoidance towards negative information, and (3) Privacy concern. Based on user responses, when mHealth apps are trying to combine text messages with app functions to deliver medical guidance, a less frequent (i.e., once or at most twice a month) and less personalized (i.e., should sound less personal but more systematic) text message is strongly preferred.

## E1. Survey Questionnaire Design

Hello, Dear users! I'm Dr. XXX from XXX Thank you for registering our diabetes management mobile application before. In order to improve the user experience, we have a few questions for you, which are expected to take you for less than 10 minutes. We will compensate you for 10 yuan after you complete the questionnaire.

**1. What is your favorite function of the blood glucose management APP? [multiple choice]**
A. Health Knowledge Function: health information
B. Diabetes Risk Assessment Function
C. Self-Tracking Function: blood glucose recording function; exercise recording function; drug recording function
D. Professional Support Function: doctor consultation function; manual personalized information guidance
E. Social Support Function: Patients' moments (patients with diabetes can view the message posted by a friend)
F. None of the above, my favorite function is
______________________

**2. Did these functions help improve my health? If so, how? If not, why? Please specify.**

______________________

**3. If we send you personalized guidance information via text messages based on your personal exercise and diet status, how would you feel? Do you think these personalized messages are helpful or not? Please specify.**

______________________

**4. Do you have any other suggestions for the function module of the blood glucose management APP? Please specify.**

______________________
**Your gender: [single choice]**
A. Male
B. female

**Your age: [single choice]**
A. Under 18
B. 18-25
C. 26-30
D. 31-40
E. 41-50
F. 51-60
G. Over 60

**Your current industry: [single choice]**
A. IT / Software and Hardware Services / E-Commerce / Internet Operations
B. Fast Moving Consumer Goods (Food / Beverage / Cosmetics)
C. Wholesale / Retail
D. Apparel / Textiles / Leather
E. Furniture / Craft / Toy
F. Education / Training / Scientific Research / Institute
G. Home Appliance
H. Communication / Telecom Operation / Network Equipment / Value-added Service
I. Manufacturing
J. Automobile and Parts
K. Catering / Entertainment / Tourism / Hospitality / Life Service
L. Office Supplies and Equipment
M. Accounting / Auditing
N. Legal
O. Bank / Insurance / Securities / Investment Bank / Risk Fund
P. Electronic Technology / Semiconductor / Integrated Circuit
Q. Instrument / Industry Automation
R. Trade / Import & Export
S. Machinery / Equipment / Heavy Industry
T. Pharmaceutical / Biotechnology/ Medical Facilities / Equipment
U. Healthcare / Nursing / Health
V. Advertising / Public Relation / Media / Art
W. Publishing / Printing / Packaging
X. Real Estate Development / Construction Engineering / Decoration / Design
Y. Property Management / Business Center
Z. Agency / Consulting / Headhunting / Certification
AA. Transportation / Logistics
BB. Aerospace / Energy / Chemical
CC. Agriculture / Fishery / Forestry
DD. Other industries

# E2. Interview Outline Design

1. **Opening**

Hello XXX, I am Dr. XXX from XXX. The purpose of this interview is to understand your attitude toward XX mobile app. This interview will take about twenty minutes. As compensation, we will pay you 20 yuan after the interview.

Your feedback and inputs are of great value to us. All your answers will be kept strictly confidential. We will not disclose your identity. All of your statements will only be used in research projects. If we want to cite any of your original words, we will use it in the form of a pseudonym.

2. **Developing**

1) Do you remember when you (refer to the registration time) registered your blood glucose mobile app?
2) What is your evaluation and impression of this blood glucose management app? What is your favorite function of diabetes management software? Why? Does these functions change your health behavior?
3) Did you receive a text message for a personalized diet and exercise guide? If so, how often?
4) Do you think these personalized text messages for health guidance helped with your health (or will help with your help if the patients have not received any)?
5) Do you think there are any disadvantages to your health caused by these text messages?
6) Would you like to cooperate with the staff to improve your eating and sports behaviors?
7) Would you like to receive a call or a text message from the company's nutritionist for active health guidance?
8) How often would you like to, if so?
9) What do you think is an acceptable personalized guidance program?
10) What do you think is the biggest pain point in the daily management of diabetes?

3. **Ending**

XXX, thank you very much for your help. I have no further questions. Your opinion is very helpful and enlightening to us. Thank you very much.